\documentclass[aps,prd,reprint,twocolumn,preprintnumbers,floatfix,nofootinbib]{revtex4-1}

\pdfoutput=1

\usepackage{graphicx}
\usepackage{epsfig}
\usepackage{amsmath}
\usepackage{amsfonts}
\usepackage{amssymb}
\usepackage{color}
\usepackage[bottom]{footmisc}
\usepackage{array,multirow,makecell}
\usepackage{slashed}
\usepackage[colorlinks,citecolor=blue]{hyperref}
\usepackage{float}
\usepackage{ulem}
\usepackage[utf8]{inputenc}
\usepackage[english]{babel}
\usepackage{url}
\usepackage{hyperref}
\usepackage{xcolor}
\usepackage{subfigure}
\usepackage{lipsum}
\usepackage[bottom]{footmisc}
\usepackage{pifont}% http://ctan.org/pkg/pifont

\usepackage[most]{tcolorbox}
\newtcbox{\mymath}[1][]{%
    nobeforeafter, math upper, tcbox raise base,
    enhanced, colframe=blue!30!black,
    colback=blue!30, boxrule=1pt,
    #1}

\usepackage{blindtext}

\def\lsim{\raise0.3ex\hbox{$\;<$\kern-0.75em\raise-1.1ex\hbox{$\sim\;$}}}
\def\gsim{\raise0.3ex\hbox{$\;>$\kern-0.75em\raise-1.1ex\hbox{$\sim\;$}}}
\newcommand{\bea}{\begin{aligned}}
\newcommand{\eea}{\end{aligned}}
\def\beq{\begin{equation}}
\def\eeq{\end{equation}}
\def\beqa{\begin{eqnarray}}
\def\eeqa{\end{eqnarray}}
\def\be{\begin{equation}}
\def\ee{\end{equation}}

\def\bse{\begin{subequations}}
\def\ese{\end{subequations}}

\def\trh{$T_{\mathrm{RH}}~$}
\def\mrm{\mathrm}
\def\tin{t_{\rm in}}
\def\tev{t_{\rm ev}}
\def\trh{T_{\mathrm{RH}}}
\def\ain{a_{\rm in}}
\def\ae{a_{\mathrm{end}}}
\def\arh{a_{\mathrm{RH}}}
\def\abh{a_{\mathrm{BH}}}

\def\rhorh{\rho_{\mathrm{RH}}}
\def\bea{\begin{eqnarray}}
\def\eea{\end{eqnarray}}
\def\rhoe{\rho_{\mathrm{end}}}

\def\aev{a_{\mathrm{ev}}}
\def\kev{k_{\rm{ev}}}
\def\Min{M_{\mathrm{in}}}
\def\Hin{H_{\rm in}}
\def\din{d_{\rm in }}
\def\kin{k_{\rm in}}
\def\rhoin{\rho_{\rm in}^{\rm BH}}
\def\nin{n_{\rm in}^{\rm BH}} 
\def\mbh{M_{\rm BH}}

\def\Tbh{T_{\rm BH}}

\newcommand{\baz}{\begin{array}{cc}}

\newcommand{\bav}{\begin{array}{cccc}}

\newcommand{\red}{\color{red}}

\begin{document}

 \title{
 Gravitational Wave Production During Reheating:\\From the Inflaton to Primordial Black Holes}

\author{Mathieu Gross}
\affiliation{Universit\'e Paris-Saclay, CNRS/IN2P3, IJCLab, 91405 Orsay, France}
\author{Essodjolo Kpatcha}
\affiliation{Departamento de Física Teórica, Universidad Autónoma de Madrid (UAM),
Campus de Cantoblanco, 28049 Madrid, Spain}
\author{Yann Mambrini}
\affiliation{Universit\'e Paris-Saclay, CNRS/IN2P3, IJCLab, 91405 Orsay, France}
\author{Maria Olalla Olea-Romacho}
\affiliation{Theoretical Particle Physics and Cosmology, King’s College London, Strand, London WC2R 2LS, United Kingdom 
}
\author{Rishav Roshan}
\affiliation{School of Physics and Astronomy, University of Southampton,
Southampton SO17 1BJ, United Kingdom}

%%%%%%%%%%%%%%%%
\begin{abstract}

We calculate the gravitational waves (GWs) produced by primordial black holes (PBHs) in the presence of the inflaton condensate in the early Universe. Combining the GW production from the evaporation process, the gravitational scattering of the inflaton itself, and the density fluctuations due to the inhomogeneous distribution of PBHs, we propose for the first time a complete coherent analysis of the spectrum, revealing three peaks, one for each source. 
Three frequency ranges ($\sim$ kHz, GHz, and PHz, respectively) are expected, each giving rise to a similar GW peak amplitude $\Omega_{\rm GW}$. We also compare our predictions with current and future GWs detection experiments.

\end{abstract}
%%%%%%%%%%%%%%%%%%
%

\maketitle

{
  \hypersetup{linkcolor=black}
  \tableofcontents
}

%%%%%%%%%%%

\section{Introduction}
\label{sec:intro}

Recently, there has been tremendous progress in the field of gravitational wave (GW) detection. The LIGO and VIRGO collaborations~\cite{LIGOScientific:2016aoc,LIGOScientific:2017vwq} have observed multiple black hole binary mergers, while the Pulsar Timing Array (PTA) collaboration~\cite{NANOGrav:2023gor,EPTA:2023fyk,Reardon:2023gzh,Xu:2023wog} has been crucial in searching for a background of stochastic GWs.

These discoveries have sparked a wave of exciting studies and have rapidly advanced the field of primordial GW research \cite{Guzzetti:2016mkm,Saikawa:2017hiv,Mazumdar:2018dfl,LISACosmologyWorkingGroup:2022jok,Gouttenoire:2019kij,Domenech:2021ztg,Roshan:2024qnv}, particularly in the context of non-standard cosmologies, where the existence and potential detection of primordial black holes (PBHs) using GW interferometry has emerged as a major focus of current research~\cite{Domenech:2024kmh, Carr:2024nlv, Green:2024bam, Flores:2024eyy, Ozsoy:2023ryl, Carr:2020gox}. The study of PBHs also provides a distinctive lens for exploring gravity, since their behavior is primarily governed by gravitational effects.

Even if a quantum theory of gravity is far from being tested in future colliders or astroparticle-dedicated experiments, the study of primordial GWs offers a unique opportunity to explore the earliest moments of the Universe, while at the same time testing (effective) quantum field theory constructions involving the graviton. This undeniable advantage of the graviton arises from its extremely weak interaction, which prohibits it from reaching thermal equilibrium with the particles in the Standard Model (SM). As a result, it preserves all the information from
the first moments of the Universe, in its spectral form.

While current experimental sensitivities, especially at the high frequencies associated with the end of inflation, have not yet reached theoretical predictions for GW signals, closing the remaining orders of magnitude may be more feasible than achieving similar results with future accelerators at comparable energy scales. Several detection proposals already exist in the literature, including laser interferometers, optically levitated sensors, and experiments based on the inverse Gertsenshtein effect, among others~\cite{Ringwald_2021,Ireland:2023avg,Roshan:2024qnv}.

It is then important to study the production of primordial GWs in a coherent framework to predict complete spectra over a wide range of frequencies.

Moreover, while different studies focus on GW
production from individual sources such as PBHs, the inflaton, or massive particles, it is essential to consider these within a broader framework to identify the complete GW spectrum observable today. In this work, we study the GW spectrum produced by the evaporation of PBHs within a generic reheating framework.

PBHs could have been formed at any stage before Big Bang nucleosynthesis (BBN), through the gravitational collapse of 
overdensities generated during inflation~\cite{Carr:1993aq, Ivanov:1994pa}, seeded by topological defects~\cite{Hawking:1987bn,Caldwell:1995fu, Jenkins:2021kcj, Blanco-Pillado:2021klh, Escriva:2022duf,Lazarides:2022spe,Lazarides:2022ezc,King:2023ayw,King:2023ztb,Bhattacharya:2023kws,Borah:2024kfn, Ghoshal:2023sfa}, or by cosmological phase transitions~\cite{Sato:1981bf, Maeda:1981gw, Liu:2021svg, Hashino:2021qoq, Baldes:2023rqv, Balaji:2024rvo, Ai:2024cka}, among others. Even if the idea of a gravitational collapsed core was proposed in 1967 by Zeldovich and Novikov~\cite{Zeldovich:1967lct}, they did not accurately addressed
the subtleties of these {\it ``cores"} in an expanding primordial background.
It was Hawking and his student Carr who finally understood and analyzed, some years later,
the formation of such objects \cite{Carr:1974nx,Carr:1975qj} and their evaporation \cite{Hawking:1975vcx,Hawking:1974rv}. The lifetime of a PBH, $\tev$, is proportional to the cube of its initial mass $i.e.~\tev \propto \Min^3$, suggesting that a heavier (lighter) PBH will have a longer (shorter) lifetime. While heavier PBHs can be good dark matter candidates provided their initial mass is above $10^{15}$ g (see~\cite{Oncins:2022ydg,Villanueva-Domingo:2021spv,Carr:2021bzv}
and references therein), lighter PBHs, with masses in the range $1~{\rm g}\lesssim\Min\lesssim 10^8$ g, can evaporate {\it before} the beginning of BBN, and produce dark matter
\cite{Bernal:2020bjf,Gondolo:2020uqv,Barman:2021ost,Barman:2022gjo,Barman:2022pdo,Chaudhuri:2023aiv}.
We should note that this behavior results from a semi-classical computation and might be modified by considering memory burden effects
\cite{Dvali:2020wft,Thoss:2024hsr,Alexandre:2024nuo,Haque:2024eyh,Chianese:2024rsn,Moursy:2024hll,Dvali:2024hsb,Barman:2024iht,Kohri:2024qpd,Zantedeschi:2024ram}
or near extremal black holes \cite{Ireland:2023avg,Arbey_2020, Cheek:2022dbx}. We have chosen not to consider these effects here. Additionally, it was shown that the evaporation of PBHs can reheat the Universe~\cite{RiajulHaque:2023cqe} or explain the origin of the matter-antimatter asymmetry~\cite{Baumann:2007yr,Hook:2014mla,Hamada:2016jnq,Barman:2024slw, Gunn:2024xaq}. 

However, at the time of PBH formation, the presence of an additional source like inflaton can also gravitationally generate radiation, dark matter, and baryogenesis. 
In fact, before reheating is complete, at a temperature $\trh$, the Universe is expected to be dominated by a homogeneous oscillating field, whose gravitational scattering through the exchange of a graviton can reheat~\cite{Clery:2022wib}, populate the dark sector~\cite{Mambrini:2021zpp,Clery:2021bwz}, generate the baryon asymmetry in a sufficient amount to be consistent with observations~\cite{Co:2022bgh},
and even produce GWs~\cite{Choi:2024bdn}.

The thermal bath itself, after the inflaton decay, can also gravitationally populate the same sectors \cite{Bernal:2018qlk}. The exchange of a graviton or the evaporation of a PBH can be regarded as \textit{two minimal} sources of production in the early Universe, as both rely solely on gravity.

Additionally, since PBH evaporation is a gravitational effect, particles are produced
democratically. However, the graviton would be the only particle that could not have entered thermal equilibrium with the other particles, thus preserving the information of the 
Hawking spectrum until now. PBH decay into gravitons has been studied in a radiation 
dominated era~\cite{Dolgov:2000ht,Anantua:2008am,Dolgov:2011cq}.
These results were obtained within instantaneous reheating
scenarios, which forbids low reheating temperature.
Recent studies~\cite{Ireland:2023avg} have considered non-standard cosmologies in which PBHs exist in a Universe dominated 
by a field with an equation of state $P_{\phi}=w_\phi\rho_{\phi}$, where $w_\phi\neq\frac13$. Compared to the well-studied case of $w=1/3$, the primary difference arises from the different dilution and redshift, which result from an expansion rate that varies with $w$.  It is impossible to disentangle a non-standard equation of state from the presence of a field that also contributes gravitationally to the production of 
the GWs. This paper aims to coherently treat the GWs produced by a population of PBHs that formed and evaporated in the presence of the inflaton field $\phi$. We then compare the GW spectrum generated by the different sources 
(inflaton, PBH evaporation, and density fluctuations in the PBH distribution), 
as a function of the potential $V(\phi)$
and the mass of the PBHs, $\Min$. It is worth noting that the current setup is also subjected to the GWs generated from quantum fluctuations during inflation \cite{Barman:2022qgt,Barman:2023ktz,Barman:2024ujh}. However, in this work, we intentionally exclude such GW production to focus specifically on the three distinct peaks arising from inflaton scattering, PBH evaporation, and density fluctuations within the PBH distribution.

The later source emerges when the black holes formed in a background dominated by an equation of state $P_\phi = w_\phi \rho_\phi$ eventually dominate the energy density of the Universe. The relative fraction of energy 
densities $\rho_{\rm PBH}/\rho_{\phi} = a^{3w_\phi}$ indicates that this scenario is possible provided that $0<w_\phi\leq 1$. Isocurvature fluctuations are inevitably 
generated once the PBHs are formed. During PBH formation, energy is transferred 
from the primordial fluid to the PBH fluid, resulting in fluctuations in the PBH fluid that are equal in magnitude but opposite in sign. This 
process keeps the total energy density constant. When PBHs come to dominate the energy density of the Universe, the initial isocurvature perturbations are converted into curvature perturbations~\cite{Papanikolaou:2020qtd, Domenech:2020ssp}, which has been recently studied in the context of a fluid with a generic equation of state parameter~\cite{Domenech:2024wao}.\footnote{We thank the authors of this paper for providing access to an early version of the article, which helped us complete this work.} These curvature perturbations, in turn, generate GWs at second order in perturbation theory~\cite{Assadullahi:2009nf, Alabidi:2013lya, Kohri:2018awv}. In this work, we will add this third source of GWs that 
must also necessarily be present in the context of PBHs formed in a $\phi$-
dominated background, should they come to dominate the energy density of the Universe.

This paper is organized as follows: after calculating the GW spectrum from PBH evaporation within a generic equation of state in Sec.~\ref{Sec:pbh}, we add the production of GWs by gravitational scattering of the inflaton itself in Sec.~\ref{sec:scatterings}. We then consider the possibility of PBH domination in Sec.~\ref{sec:pbhdomination}, adding the contribution due to the inhomogeneities in their distribution. Finally, we summarize our work before concluding.
Our main results are nicely illustrated in Fig.~(\ref{Fig:master1}), which show the GW spectrum including all the processes mentioned above.

\section{Gravitational waves from PBH evaporation} 
\label{Sec:pbh}

In this section, we discuss a scenario in which the Universe transitions from the $w$-dominated epoch to radiation domination. In this case, PBHs will play a subdominant role in the energy budget of the Universe, from their production to their evaporation. Nevertheless, they should still contribute to the gravitational wave energy density through their evaporation process.

\subsection{Generalities}

Primordial black holes, formed in the earliest stages of the Universe, are a potential source of primordial gravitational waves through their Hawking radiation

\beq
\frac{d \mbh}{dt}=-\epsilon \frac{M_P^4}{\mbh^2}\,,
\label{Eq:hawking}
\eeq
with 
$\epsilon =\frac{27}{4}\frac{\pi g_{\star}(\Tbh)}{480}$,
$g_{\ast}(\Tbh)$ is the number of degrees of freedom 
associated with the PBH temperature, and 
$M_P = 1/\sqrt{8\pi G} \simeq 2.4 \times 10^{18} \, \rm{GeV}$ is the reduced Planck mass. 
The factor ${27}/{4}$~\cite{MacGibbon:1990zk} accounts for the greybody factor.\footnote{Which can be more complex depending on the context \cite{Auffinger:2020afu,Masina:2021zpu,Cheek:2021odj}.}  Note that we neglected the accretion effect.

Solving Eq.~(\ref{Eq:hawking}) leads to the time evolution of the PBH mass

\beq
\mbh(t)=\Min\left(1-\frac{t-\tin}{\tev}\right)^\frac13,
\label{Eq:mbht}
\eeq
where $t_{\rm in}$ is the time at formation, and $\tev$ is the evaporation time, i.e.

\beq
\tev =\frac{\Min^3}{3\epsilon M_P^4}
\simeq 2.4\times 10^{-28} \left(\frac{\Min}{1~\rm{g}}\right)^3~\rm{s}\,.
\label{Eq:tev}
\eeq
PBHs with masses $\lesssim 10^8$g decay {\it before}
the beginning of the nucleosynthesis process, and are not constrained by CMB data. The initial PBH mass $M_{\rm in}$ is bounded by the size of the horizon at the end of inflation,
%%%%%%%%%%%%%%%%%%%%%%%%%
\bea
\Min \gtrsim H_{\rm end}^{-3} \rho_{\rm end} \sim \frac{M_P^3}{\sqrt{\rho_{\rm end}}}\simeq 1\mrm{g}\,.
\eea
Taking into account the above constraints, we will restrict our analysis to the following PBH mass range: 
%%%%
\bea
1 \mrm{g} \lesssim \Min \lesssim 10^8 \mrm{g}.
\eea

It was shown 
in \cite{RiajulHaque:2023cqe, Haque:2024eyh,Haque:2024cdh}
that PBHs can reheat the Universe with their Hawking radiation. Even if this is not the case,
there is the possibility that the graviton emission
from the evaporation can still contribute significantly to present
stochastic GW spectrum.
From the Hawking temperature \cite{Hawking:1975vcx,Hawking:1974rv}

\beq
\Tbh = \frac{M_P^2}{\mbh}\,,
\eeq
 one can deduce the spectrum of outgoing particles of species $i$, 

\beq
\frac{dN_i}{dt}=\frac{27}{4}\pi R_S^2\frac{g_i}{(2 \pi)^3}\frac{d^3k}{e^{\frac{k}{\Tbh}}\pm 1},
\label{Eq:distribution}
\eeq
where $k$ represents the particles momentum, $g_i$ denotes their internal degrees of freedom, and $R_S = \frac{\mbh}{4 \pi M_P^2}$ is the Schwarzschild radius. In the case of gravitons,
with two degrees of freedom, the energy spectrum $\rho_{\rm GW}^{\rm BH}$
becomes

\beq
\frac{d\rho^{\rm BH}_{\rm GW}}{dtdk}=n_{\rm BH}(t)k\frac{dN_{i}}{dtdk}
\,,
\label{Eq:pbhgwspectrum}
\eeq

where $n_{\rm BH}(t)$ is the PBH number density.

\subsection{Gravitational wave spectrum from PBH decay}

To compute
the GW spectrum at the evaporation time, 
we integrate Eq.~(\ref{Eq:pbhgwspectrum})
over the lifetime of the PBH. We  express all the dynamical quantities as a function of the scale factor $a$. In particular, the PBH number density reads

\beq
n_{\rm BH}(t)= \nin \left(\frac{\ain}{a}\right)^3\,,
\eeq
where $\nin$ is the PBH number density at the formation time, corresponding to the scale factor $\ain$. The initial PBH mass can be written as a fraction of the horizon size at formation as
\beq
\Min= \frac{4 \pi}{3} \gamma\,\frac{ \rho_{\rm tot} (\ain)}{\, \Hin^3}=4 \pi \gamma \frac{M_P^2}{\Hin} \,,
\label{Eq:min}
\eeq
where  $\gamma$ represents the efficiency factor for collapse, and $\gamma\sim 0.2$ (for radiation domination) \cite{Carr:1974nx}, or $\gamma \sim w^\frac32$ \cite{Villanueva-Domingo:2021spv} in a Universe dominated by a general equation of state $P=w\rho$. From the former expression, we have
\beq
\nin=\frac{\rhoin}{\Min}=48 \beta \pi^2 \gamma^2 \frac{M_P^6}{\Min^3}\,,
\label{Eq:nin}
\eeq
where $\rhoin=\beta \rho_{\rm tot}(\ain)$.

The PBH mass
$\mbh(t)$ can also be written as function of the scale parameter, expressing
$t=f(a)$ in Eq.~(\ref{Eq:mbht}), we obtain

\bea
&&
\mbh(a)\simeq\Min\left[1-\left(\frac{a}{\aev}\right)^{\frac{3(1+w)}{2}}\right]^\frac13
\nonumber
\\
&&
=\Min\left[1-\alpha_{\rm BH}\left(\frac{a}{\ain}\right)^{\frac{3(1+w)}{2}}\right]^\frac13 \, ,
\label{Eq:mbha}
\eea
with 
\beq
\alpha_{\rm BH}= \left(\frac{\ain}{\aev}\right)^\frac{3(1+w)}{2}
=\frac{27}{4}\frac{g_*(\Tbh)}{960 \gamma(1+w)}\frac{M_P^2}{\Min^2}\,,
\eeq
where we used Eqs.~(\ref{Eq:tev}) and (\ref{Eq:min}) together with the relation $t_{\rm in}=\frac{2}{3(1+w)}H_{\rm in}^{-1}$ to obtain

\bea
&&
\frac{\aev}{\ain}=\left(\frac{\tev}{\tin}\right)^\frac{2}{3(1+w)}
=\left(\frac{3(1+w)}{2}\right)^\frac{2}{3(1+w)}\left(\tev \Hin\right)^\frac{2}{3(1+w)}
\nonumber
\\
&&
=\left(\frac{4}{27}\frac{960 \, \gamma\,(1+w)}{\,g_*(\Tbh)}\,
\frac{\Min^2}{M_P^2}\right)^\frac{2}{3(1+w)}
\, .
\label{Eq:aevain}
\eea
We considered $a\gg\ain$ ($t\gg \tin$). The mode $k$ can also be expressed as function of $a$ and the present mode $k_0$,  
\beq
k=k_0\left(\frac{a}{a_0}\right)^{-1}\,.
\eeq

Finally, writing $dt=\frac{da}{Ha}$, with 
$H(a)=\Hin\left(\ain/a\right)^\frac{3(1+w)}{2}$, we can integrate
 Eq.~(\ref{Eq:pbhgwspectrum}) over the entire evaporation time to obtain

\bea
&&
\frac{d \rho_{\rm GW}^{\rm BH}}{d\ln k_0}=\frac{27\nin k_0^4}{64 \pi^3M_P^4}\int^{\aev}_{\ain}\frac{\mbh^2(a)}{{e^{\frac{k_0\mbh}{M_P^2}\frac{a_0}{a}}-1}}
\left(\frac{\ain}{a}\right)^3\frac{da}{Ha}
\label{Eq:drhoGWBH}
\\
&&
=\frac{81 \beta \gamma k_0^4}{16\pi^2}\int_{\ain}^{\aev}
\frac{\left(1-\alpha_{\rm BH}\left(\frac{a}{\ain}\right)^\frac{3+3w}{2}\right)^\frac23}{e^{\frac{k_0\mbh}{M_P^2}\frac{a_0}{a}}-1}
\left(\frac{\ain}{a}\right)^\frac{3-3w}{2}
\frac{da}{a},
\nonumber
\eea
where we expressed the spectrum as a function of the 
mode $k_0$. The redshift factor of $d\rho_{\rm GW}^{\rm BH}|_0\propto (\frac{\aev}{a_0})^4$, incorporating the evolution from the evaporation time to the present day, cancels out with that of $k^4$.

In the limit in which $\mbh\sim \Min$, after appropriate changes of variables
between $\ain$ and $\aev$, one obtains

\bea
&&
\frac{d \rho_{\rm GW}^{\rm BH}}{d\ln k_0}
\simeq 
\frac{81 \beta \gamma k_0^{\frac{5+3w}{2}}}{16\pi^2}
\left(\frac{\ain}{a_0}\right)^\frac{3-3w}{2}
\left(\frac{M_P^2}{\Min}\right)^{\frac{3-3w}{2}}
\nonumber
\\
&&
\times
\int_{\frac{k_0\Min}{M_P^2}\frac{a_0}{\aev}}^{\frac{k_0\Min}{M_P^2}\frac{a_0}{\ain}}
\frac{X^{\frac{1-3w}{2}}}{e^X-1}\,dX
\label{Eq:spectrumaev}
\\
&&
=
\frac{81 \beta \gamma k_0^{\frac{5+3w}{2}}}{16\pi^2}
\left(\frac{\ain}{a_0}\right)^\frac{3-3w}{2}
\left(\frac{M_P^2}{\Min}\right)^{\frac{3-3w}{2}}
I(k_0,\Min,w,\trh)\,,
\nonumber
\eea
where $\trh$ is the reheating temperature coming from an unspecified process. Therefore, we treat it as a free parameter in our analysis.
We defined

\beq
I(k_0,\Min,w,\trh)=\int_{\frac{k_0\Min}{M_P^2}\frac{a_0}{\aev}}^{\frac{k_0\Min}{M_P^2}\frac{a_0}{\ain}}
\frac{X^{\frac{1-3w}{2}}}{e^X-1}\,dX\,,
\label{Eq:i0}
\eeq
and assumed from Eq.~(\ref{Eq:mbht}), that $\mbh \simeq \Min$ during 
the whole process of evaporation, or, in other words, that the mass loss through Hawking radiation is negligible compared to the PBH mass until the complete evaporation. We compute the integral $I(k_0,\Min,w,\trh)$ numerically.
Next, we express $\frac{a_0}{\aev}$, which appears in the integration limits of Eq.~(\ref{Eq:spectrumaev}), as a function of $\Min$ by using

\bea
&&
\frac{\aev}{\arh}=\left(\frac{\tev}{t_{\rm RH}}\right)^\frac{2}{3(1+w)}
=\left(\frac{3 (1+w)H(\arh)\tev}{2}\right)^\frac{2}{3(1+w)}
\nonumber
\\
&&
=\left(\frac{320\sqrt{3}(1+w)\sqrt{\rhorh}\Min^3}{27 g_*\pi M_P^5}\right)^\frac{2}{3(1+w)}\,,
\label{Eq:aevarh}
\eea
where $\rhorh=\frac{g(\trh)\pi}{30}\trh^4=\alpha\trh^4$\,, which gives

\bea
\frac{\aev}{a_0}=\frac{\aev}{\arh}\frac{\arh}{a_0}=
&&
\left(\frac{g_0}{g_{RH}}\right)^\frac13\frac{T_0}{\trh^\frac{3w-1}{3w+3}}
\label{Eq:aevova0}
\\
&&
\times\left(\frac{320(1+w)\sqrt{3 \alpha}\Min^3}{27 g_*(\Tbh)\pi M_P^5}\right)^\frac{2}{3(1+w)}\,,
\nonumber
\eea
where we used $\frac{\arh}{a_0}=\left(\frac{g_0}{g_{\rm RH}}\right)^\frac13\frac{T_0}{\trh}$.
To obtain Eq.~(\ref{Eq:spectrumaev}), we also wrote 

\beq
k_0\frac{a_0}{a}=k_0\frac{a_0}{\ain}\frac{\ain}{a}
 \,,
 \eeq
with
\bea
&&
\frac{\ain}{a_0}=\frac{\ain}{\aev}\frac{\aev}{\arh}\frac{\arh}{a_0}
\label{Eq:aiova0}
\\
&&
=
\left(\frac{g_0}{g_{RH}}\right)^\frac13\frac{T_0}{\trh^\frac{3w-1}{3w+3}}
\left(\frac{\alpha}{3}\right)^\frac{1}{3+3w}
\left(\frac{\Min}{4 \pi \gamma M_P^3}\right)^\frac{2}{3+3w}
\,,
\nonumber
\eea
and we combined Eqs.~(\ref{Eq:aevain}) and (\ref{Eq:aevarh}).

\if{
\begin{figure}[H]
    \centering
    \includegraphics[scale=0.8]{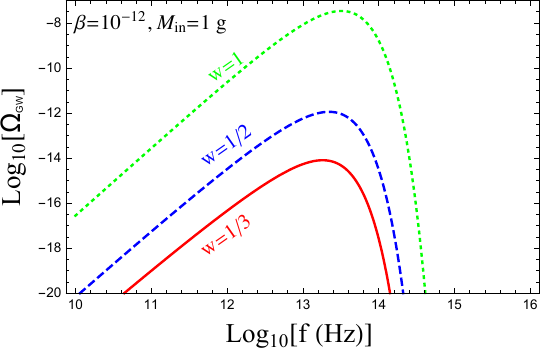}
     \caption{\it Examples of gravitational wave spectrum generated by PBH decay for $\beta=10^{-12}$, $\trh=10^{10}$ GeV and $\Min=1$ g, with
     different values of equation of state: $w=1/3$ (full red), 1/2 (dashed blue) and 1 (dotted green). The result for different values
     of $\beta$ can be simpy rescaled as $\Omega_{\rm GW}^{\rm BH}\propto \beta$.{\red \bf [Can someone check the plot with BlackHawk?]}}
    \label{Fig:gwfrompbh}
\end{figure}
}
\fi

We show in Fig.~(\ref{Fig:gwfrompbh}) the spectrum for different  
equations of state ($w=0$, $\frac13$, $\frac12$ and $1$)
for $\beta=10^{-12}$, $\Min=1 \, \rm{g}$,
and two reheating temperatures, $\trh=10^{10}$ GeV (dashed), 
and $\trh=10^{13}$ GeV (solid).\footnote{Note that for
$\beta=10^{-12}$, with reheating temperature $\trh \lesssim 10^{10}$ GeV, PBHs with mass $M_{\in}= 1 \, \rm{g}$ would dominate the reheating process over the inflaton (see Fig.~(6) of \cite{RiajulHaque:2023cqe}),
and $\trh$ would not be independent of $\Min$, which is the subject of the following section.} Although no PBHs are expected to form at $w=0$~\cite{Escriv__2021}, arbitrarily setting $\gamma = 1$ and $w=0$ allows us to explore the potential contributions of a PBH population in this scenario. This approach represents a situation where PBHs are formed by an unspecified mechanism, to which we remain agnostic.
We notice that for $w>1/3$, the peak amplitude decreases when $T_{\rm RH}$ increases, and viceversa for $w<1/3$. 
The spectrum was obtained using the state-of-the-art tool, the \texttt{BlackHawk} package~\cite{Arbey_Auffinger_2019,Arbey_Auffinger_2021}. 
\texttt{BlackHawk} allows to simulate the evolution of a PBH population beyond the blackbody approximation, incorporating tabulated greybody factors. The result is the GW spectrum generated by the instantaneous emissions of the PBHs, which contribute to the complete graviton spectra.

To understand the shape of the spectrum, we can develop
$e^X\simeq 1+X$ for 
$X\ll1$ in Eq.~(\ref{Eq:i0}),\footnote{ 
Note that the analytical ansatz for this limit differs slightly with the slope obtained in our numerical simulation due to greybody factor effects when $\frac{\Min}{T_{\rm BH}}\ll 1$ \cite{Cheek:2021odj}.} and account for an exponential suppression when $X \gg 1$, which corresponds to $k(\aev)=\kev \gg \Tbh=\frac{M_P^2}{\Min}$. In this limit, Eq.~(\ref{Eq:spectrumaev}) yields

\bea
&&
\left. \Omega_{\rm GW}^{\rm BH}\right|_{\kev\ll \Tbh}\simeq
\frac{\beta}{10^{-10}}\left(\frac{f_0}{10^{10}~{\rm Hz}}\right)^3
\times
\label{Eq:Omegasmallk}
\\
&&
\begin{cases}
5.1\times 10^{-21}~
\left(\frac{1~\rm{g}}{\Min}\right)^\frac{3w+1}{3w+3}
\left(\frac{10^{10}~\rm{GeV}}{\trh}\right)^\frac{3w-1}{3w+3}
\text{for}~w<\frac13
\nonumber
\\
2.8\times 10^{-20}~
\left(\frac{1~\rm{g}}{\Min}\right)^\frac12
\left[1+0.08\log\left(\frac{\Min}{1~\rm{g}}\right)\right]
\text{for}~w=\frac13
\nonumber
\\
2.8\times 10^{-15}~
\left(\frac{1~\rm{g}}{\Min}\right)^\frac{1-w}{1+w}
\left(\frac{10^{10}~\rm{GeV}}{\trh}\right)^\frac{3w-1}{3w+3}
\text{for}~w>\frac13
\,,
\end{cases}
\eea
where we defined the present GW spectral density produced by a source $i$ as
\beq
\Omega^i_{\rm GW}=\frac{1}{\rho_c} \left.\frac{d \rho_{\rm GW}^{i}}{d\ln k}\right|_{a_0}\,,
\eeq
with $\rho_c$ being the present critical density $\rho_c=3 H_0^2M_P^2$. This approximation effectively describes the constant slope of the spectrum at low frequencies. In the next section, we discuss the GW amplitude at the peak, which is crucial for comparison with the sensitivity of future experiments.

\begin{figure}[H]
    \centering
    \includegraphics[width=1\linewidth]{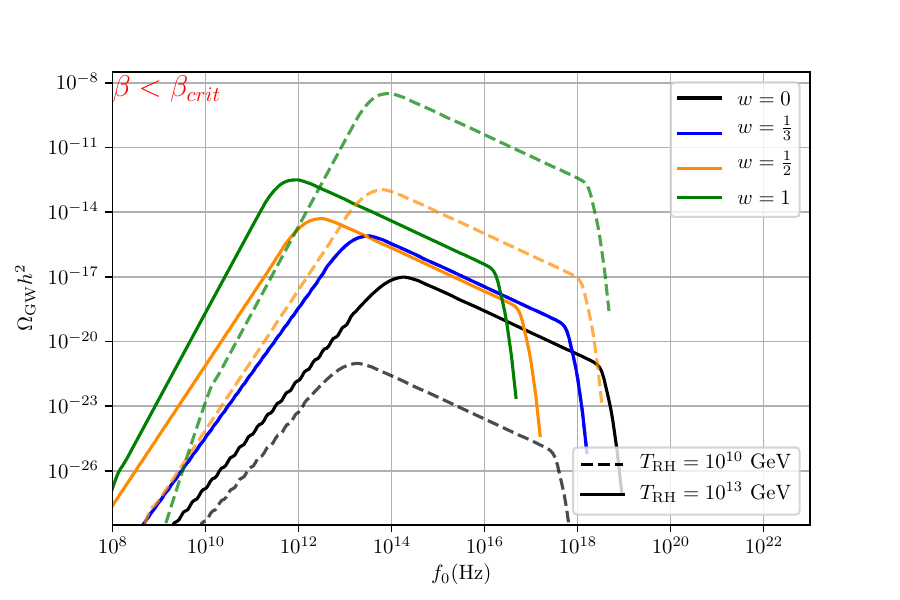}
    \caption{Gravitational wave spectrum generated by PBH decay for $\Min=1$ g and $\beta=10^{-12}$, for different values of $w$ and the reheating temperature, $\trh=10^{13}$ GeV (solid lines) and $\trh=10^{10}$ GeV (dashed lines). 
    For $w=\frac{1}{3}$, the spectrum is independent of $\trh$, and both curves overlap.}
    \label{Fig:gwfrompbh}
\end{figure}

\if
{
\begin{figure}[H]
    \centering
    \includegraphics[width=\linewidth]{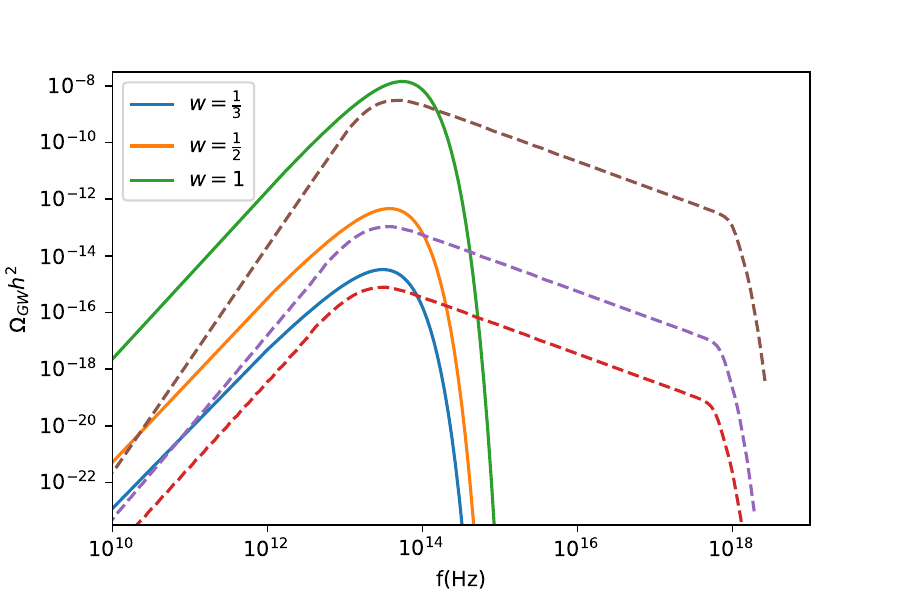}
    \caption{gravitationnal wave spectrum from PBH decay for $\beta = 10^{-12}$, $M_{in} = 1g$ and $T_{rh}=10^{10}$GeV, the dashed line corresponds to numerical simulation while the hick line are the analytical approximation from Eq.~(\ref{Eq:spectrumaev}). }
    \label{fig:enter-label}
\end{figure}
}
\fi

\subsection{Peak gravitational wave amplitude from PBH evaporation}

The approximations in Eq.~(\ref{Eq:Omegasmallk})
are in agreement with
the numerical results shown in Fig.~(\ref{Fig:gwfrompbh}): below the value at the peak frequency $f^{\rm peak}$,
when the exponential distribution begins to dominate the spectrum, we recover the slope $\Omega_{\rm GW}^{\rm BH} \propto f_0^3$.  
The GW energy density spectrum reaches its maximum for\footnote{The maximum of the function $f(X)=\frac{X^3}{e^X-1}$ is reached for $X^{\rm peak}\simeq 2.8$.}

\bea
k_0^{\rm peak} \simeq
&&
2.8 \frac{M_P^2}{\Min}\frac{\aev}{a_0}\simeq 2.8\,
T_0\frac{g_0^\frac13}{g_{RH}^\frac13}\left(\frac{\Min}{M_P}\right)^\frac{1-w}{1+w}
\label{Eq:k0peak}
\\
&&
\times
\left(\frac{M_P}{\trh}\right)^\frac{3w-1}{3w+3}
\left(\frac{320\sqrt{3}(1+w)\sqrt{\alpha}}{27 g_*(\Tbh) \pi}\right)^\frac{2}{3+3w}\,,
\nonumber
\eea
which gives

\bea
&&
f^{\rm peak}_{w=\frac13} \simeq  1.75\times 10^{13}\sqrt{\frac{\Min}{1\,\rm{g}}}\,{\rm Hz}\,,
\label{Eq:fpeakbh}
\\
&&
f^{\rm peak}_{w=\frac12} \simeq  2\times 10^{13} \left(\frac{\Min}{1\,{\rm g}}\right)^\frac13\left(\frac{10^{10}\,{\rm GeV}}{\trh}\right)^\frac19\, {\rm Hz}\,,
\nonumber
\\
&&
f^{\rm peak}_{w=1} \simeq  2.9\times 10^{13}\left(\frac{10^{10}\,{\rm GeV}}{\trh}\right)^\frac13\, {\rm Hz}\,,
\nonumber
\eea
for different choices of $w$. For $\trh = 10^{13}, \, 10^{10}$ GeV, these values are in agreement with the peak frequencies displayed in Fig.~(\ref{Fig:gwfrompbh}). 

\begin{figure}[H]
    \centering
    \includegraphics[scale=0.8]{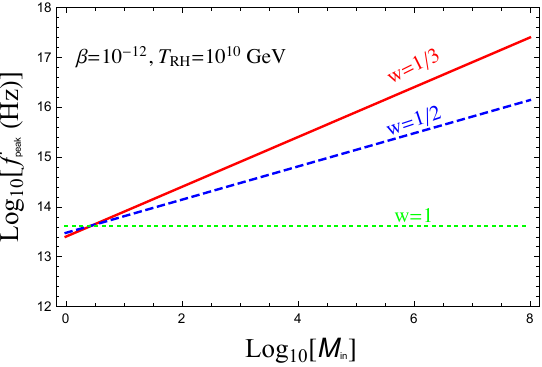}
     \caption{Present peak frequency of the gravitational wave density as a function of $\Min$,
     for $\trh=10^{10}$ GeV
     and different values of $w$, $w=\frac13$,
     $\frac12$ and $1$.}
    \label{Fig:fpeakpbh}
\end{figure}

Note that $f^{\rm peak}$ is independent of $\trh$ for $w=1/3$,
which is expected since a Universe dominated by 
a radiation--type field before the completion of reheating 
is indistinguishable from one dominated by pure radiation. Thus, for $w=1/3$, the
reheating process plays no role in the redshifting of the frequency, 
whereas $f^{\rm peak}_{w=\frac12}\propto \trh^{-\frac19}$ and $f^{\rm peak}_{w=1}\propto \trh^{-\frac13}$. We show in Fig.~(\ref{Fig:fpeakpbh}) the present peak frequency as a function of $\Min$ for different values of the
equation of state $w=\frac12$, $\frac13$ and $1$. We remark that the frequency does not depend on $\Min$ for $w=1$
because the redshift factor goes as $\frac{\aev}{a_0}\propto \Min$ (see Eq.~(\ref{Eq:aevarh})), so the mass drops out from
the frequency at the production time $\propto \frac{M_P^2}{\Min}$.

\begin{figure*}[ht!]
    \centering
    \subfigure{\includegraphics[scale=0.8]{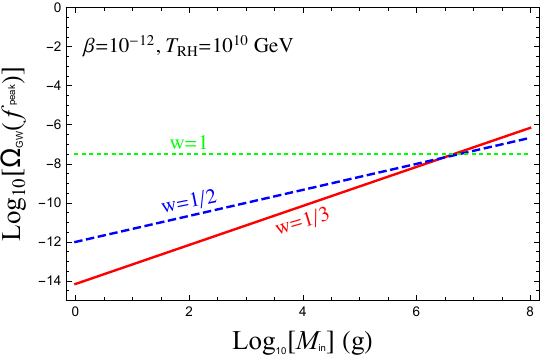}}
    \hspace{0.5cm}
\subfigure{\includegraphics[scale=0.8]{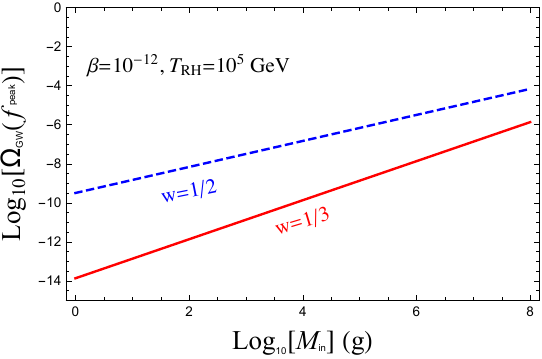}}
     \caption{Gravitational wave density parameter $\Omega_{\rm GW}^{\rm BH}$
     at the present peak frequency $f_0^{\rm peak}$ as function of the PBH mass
     $\Min$ for $\trh=10^{10}$ GeV (left) and $\trh=10^{5}$ GeV (right) and different values of $w=\frac13$, $\frac12$ and $1$. The result for different values of $\beta$ is obtained by a simple rescaling. On the right plot $w=1$ is not present due to the PBH reheating effect for this case (see section IV for details).}
    \label{Fig:Omegapeakpbh}
\end{figure*}

From the detection perspective, it is interesting to compute the peak GW density, $\Omega_{\rm GW}^{\rm BH}(f_0^{\rm peak})$ as a function 
of $\Min$ and $\trh$ for a given PBH initial density $\beta$. The result is shown in Fig.~(\ref{Fig:Omegapeakpbh})
for different values of $w$. Here, we have fixed $\beta=10^{-12}$  and set the reheating temperature to $\trh=10^{10}$ GeV (left) and $\trh=10^5$ GeV (right). 

The behavior of $\Omega_{\rm GW}^{\rm BH}(f_0^{\rm peak})$ can be obtained
by substituting Eq.~(\ref{Eq:k0peak}) into Eq.~(\ref{Eq:Omegasmallk}), which gives

\beq
\Omega_{\rm GW}^{\rm BH}(f_0^{\rm peak})\,\propto\,\beta\,\frac{\Min^{\frac{2-2w}{1+w}}}{\trh^{\frac{12w-4}{3w+3}}}\,.
\label{Eq:omegapeakpbh}
\eeq

The peak GW amplitude decreases with lower reheating temperatures when $w < \frac{1}{3}$, remains constant for $w = \frac{1}{3}$, and increases with lower $T_{\text{RH}}$ when $w > \frac{1}{3}$.
For $w>\frac13$, the GW density $\propto a^{-4}$ redshifts less than the background between $\aev$
and $\arh$. As a result, its relative abundance at the end of reheating is increased if reheating lasts longer, corresponding to lower reheating temperatures.

We plotted the condition Eq.~(\ref{Eq:omegapeakpbh}) in Fig.~(\ref{Fig:Omegapeakpbh})
for $\trh=10^{10}$ GeV (left) and $10^5$ GeV (right). Our analytical estimation for $\Omega_{\rm GW}^{\rm BH} (f_0^{\rm peak})$ 
matches the numerical prediction obtained by \texttt{BlackHawk} in Fig.~(\ref{Fig:gwfrompbh}). 
We note that, while the peak density does not depend on $\Min$ in the 
kination regime ($w = 1$), it depends on the reheating temperature, i.e.~$\Omega_{\rm GW}^{\rm BH}(f_0^{\rm peak})\big|_{w=1} \propto \trh^{\frac{4}{3}}$. This implies that a change in $\trh$ by 3 orders of magnitude results in an increase in the GW density by a factor of $10^4$, which aligns with what we observe in Fig.~(\ref{Fig:Omegapeakpbh}).
Physically, this effect is coming from the dependence of the peak frequency on 
$\trh$ for large values of $w$ because a larger pressure ($P=\rho$ in the kination case)
slows down the expansion rate, and, thus, the dilution of the GW spectrum. 

Notably, the GW spectrum shows a second breaking for frequencies $f_0>f_0^{\rm peak}$, as seen in Fig.~(\ref{Fig:gwfrompbh}). Indeed, for this regime of frequencies,

the PBH decay cannot be considered instantaneous, although we can assume $M_{\rm BH}=\Min$ throughout most of the decay process. 
At $a\simeq \aev$, the PBHs lose their mass
rapidly, following Eq.~(\ref{Eq:mbha}). As a consequence, there is a sudden increase
of $\Tbh = \frac{M_P^2}{M(a)}$, allowing for larger momentum modes to be produced.
The GW spectrum at large $k$ is then enhanced
when one takes 
into account the exact evolution of $M_{\rm BH}(a)$ in Eq.~(\ref{Eq:drhoGWBH}), which what we observe
before the exponential suppression of the spectra in Fig.~(\ref{Fig:gwfrompbh}).
This corresponds to a shift of the peak toward higher values following $f_0^{\rm peak}\propto T_{\rm BH}$, but with a decreased amplitude due to the PBH dilution with time.

Since in this section we focus specifically on the case in which the energy density of the inflaton field $\phi$ dominates the energy budget during reheating, the GW spectrum generated by inflaton scattering also becomes significant and must be compared to the spectrum produced by PBH evaporation before reheating completes. The presence of the inflaton contributes to GW production, and that cannot be neglected in non-standard cosmologies, a point we explore in detail in the following section.

\section{Gravitational waves from inflaton scattering}
\label{sec:scatterings}

\subsection{Generalities}

After inflation, the Universe goes through a phase of
oscillation of the inflaton $\phi$, following an equation of state 
$P_\phi =w_\phi \rho_\phi$, with

\beq
w_\phi =\frac{k-2}{k+2}\,,
\label{Eq:wphi}
\eeq
for a potential \cite{Garcia:2020eof,Garcia:2020wiy}
\beq
V(\phi)=\lambda M_P^4 \left(\frac{\phi}{M_P}\right)^k\,.
\label{Eq:inflatonpotential}
\eeq

This potential\footnote{Obviously, the $k$ in Eq.~(\ref{Eq:wphi}) has nothing to do with the momentum $k$ of Eq.~(\ref{Eq:distribution}), and we do not think confusions are possible in our expressions throughout the paper.} can be seen as the limit in which $\phi \ll M_P$ of, for instance, 
a Starobinsky potential \cite{Starobinsky:1980te} of the form
\beq
V(\phi)=\frac34 m_\phi^2M_P^2\left(1-e^{-\sqrt{\frac23}\frac{\phi}{M_P}}\right)^2\,,
\eeq
or $\alpha$-attractor T-models~\cite{Kallosh:2013hoa}

\beq
V(\phi)=\lambda M_P^4\left[\sqrt{6} \tanh \left(\frac{\phi}{\sqrt{6}M_P}\right)\right]^k\,.
\eeq

The condition of a dominant $\phi$-energy density with $\rho_\phi \sim$ constant is expressed in the Boltzmann equation

\beq
\dot \rho_\phi +3(1+w_\phi)H \rho_\phi\simeq 0\,,
\label{Eq:diffrhophi}
\eeq
which implies
\beq
\rho_\phi = V(\phi_0)=\rhoe\left(\frac{\ae}{a}\right)^{3(1+w_\phi)}=
\rhoe\left(\frac{\ae}{a}\right)^{\frac{6k}{k+2}}\,,
\label{Eq:rhophi}
\eeq
with $\rhoe$ being the inflaton energy density at the end of inflation.
For our numerical analysis, we used the following set of parameters

\begin{align}
    &\lambda = \frac{18\pi^{2}A_{S}}{6^{k/2}N_{*}},
    \nonumber
    \\
    &\phi_{\rm end} = \sqrt{\frac{3}{8}}M_{p}\ln\left[\frac{1}{2}+\frac{k}{3}(k+\sqrt{k^{2}+3})\right]
    \\
    &m_{\phi_{ \rm end}} = \sqrt{k(k-1)\lambda}\left(\frac{\phi_{\rm end}}{M_{p}}\right)^{\frac{k-2}{2}}M_ {p},
    \nonumber
    \\
    &\rho_{\phi}\propto a^{-\frac{6k}{k+2}}, \hspace{0.5cm}m_{\phi} \propto a^{-3\frac{k-2}{k+2}}, \hspace{0.5cm}H_{ \rm end} \sim 10^{-6}M_{p}\,,
    \nonumber
\end{align}
where $A_{S}= e^{3.044}/10^{10}$ is the amplitude of the scalar power spectrum, and we chose $N_{*}=55$. We highlight that for the specific case $w=1$, we took the limit $k\rightarrow \infty$.

During this oscillatory phase, the energy stored in the inflaton condensate is transferred to quantum states potentially relevant for processes such as reheating, dark matter~\cite{Choi:2024bdn}, or GW production~\cite{Bernal:2023wus,Barman:2023rpg,Barman:2023ymn}. This energy transfer can be interpreted as a result of inflaton scattering or decay processes, which may involve physics beyond the SM.
However, recent studies suggest that even minimal gravitational interactions, albeit Planck-suppressed, may be sufficient to complete reheating~\cite{Clery:2021bwz,Clery:2022wib,Haque:2022kez}, populate the dark sector~\cite{Bernal:2018qlk,Mambrini:2021zpp}, drive leptogenesis~\cite{Barman:2022qgt,Barman:2024kyw}, and produce potentially observable gravitational waves~\cite{Choi:2024ilx}.

Interpreting the background field as an oscillating inflaton impacts the GW spectrum in three key ways. First, the equation of state $P = w_\phi \rho$ in Eqs.~(\ref{Eq:mbht}) and (\ref{Eq:spectrumaev}) is determined by the shape of the inflaton potential $V(\phi)$, as described in Eq.~(\ref{Eq:wphi}). Second, the reheating temperature is set by the moment when the energy density transferred to radiation, $\rho_R$, equals the inflaton energy density $\rho_\phi$ according to the evolution:

\begin{equation}
\dot{\rho}_R + 4 H \rho_R = (1 + w_\phi) \Gamma_\phi \rho_\phi.
\label{Eq:diffrhor}
\end{equation}

\begin{figure}[H]
    \centering
    \includegraphics[scale=0.8]{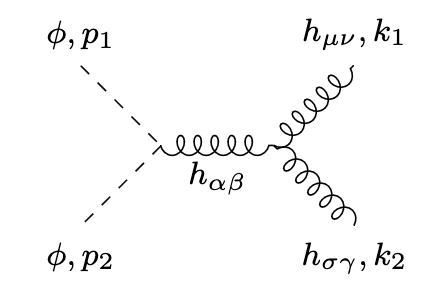}
     \caption{Example of a purely gravitational process contributing to the gravitational wave background through inflaton scattering.}
    \label{Fig:Feynmaninflaton}
\end{figure}

 This equality occurs at $\arh$ defined by $\rho_\phi(\arh)=\rhorh=\alpha \trh^4$. Solving Eqs.~(\ref{Eq:diffrhor}) and (\ref{Eq:diffrhophi}),
we obtain~\cite{Garcia:2020eof,Garcia:2020wiy}

\beq
\rho_R=\rhorh\left(\frac{\arh}{a}\right)^{\frac32(1+3w_\phi)}=\rhorh\left(\frac{\arh}{a}\right)^\frac{6k-6}{k+2}.
\eeq
Finally, in the presence of the inflaton field, the GW spectrum must necessarily account for gravitational scattering processes, such as the one shown in Fig.~(\ref{Fig:Feynmaninflaton}) \cite{Choi:2024ilx}, in addition to any other relevant GW processes.
Therefore, this contribution should be added to the PBH evaporation spectrum obtained in Eq.~(\ref{Eq:spectrumaev}) and Fig.~(\ref{Fig:gwfrompbh}).

In the context of minimal production, the GW energy density obeys the following Boltzmann equation

\beq
\dot \rho_{\rm GW}^\phi+4 H \rho_{\rm GW}^\phi=(1+w_\phi)\Gamma_\phi^{\rm GW}\rho_\phi
\eeq
or, equivalently,

\begin{equation}
    \frac{1}{a^{4}}\frac{d a^{4}\rho_{\rm GW}^\phi}{dt} = R^\phi_{\rm GW} \,\Rightarrow\,
    \frac{d a^{4}\rho_{\rm GW}^\phi}{da}=\frac{a^3}{H}R^\phi_{\rm GW}\,,
    \label{Eq:rhogwphi}
\end{equation}
where $R^\phi_{\rm GW}$ is the rate of GW production due to inflaton scatterings (energy transferred per unit of time and unit of volume), given by~\cite{Clery:2021bwz,Choi:2024ilx}
\begin{equation}
    R^\phi_{\rm GW} = 2 \times \frac{\rho_{\phi}^{2}\omega}{4\pi M_{P}^{4}}\Sigma^{k}\,,
\end{equation}
with
\beq
    \Sigma^{k} = \sum_{n = 1}^{\infty} n\left| \left(\mathcal{P}^{k}\right)_{n}\right|^{2}\,.
\label{Eq:sigmak}
\eeq
Here we accounted for the production of two gravitons per scattering. Note also that we used the usual ``envelope approximation" for the inflaton evolution $\phi(t) = \phi_{0}(t)\mathcal{P}(t)$, where $\mathcal{P}(t)$ encodes all the anharmonicity of the inflaton rolling. It was shown in refs.~\cite{Clery:2021bwz,Clery:2022wib,Choi:2024ilx} that the sum in Eq.~(\ref{Eq:sigmak}) is largely dominated by the first mode (corresponding to $n=2$ in our convention), and the Fourier coefficients decrease rapidly with $n$.
Therefore, we will consider only the first mode as an approximation for the remainder of our analysis. We then get the Fourier frequency for the dominant mode \cite{Garcia:2020eof,Garcia:2020wiy,Clery:2021bwz}:

\begin{equation}
    \omega = m_{\phi}\sqrt{\frac{\pi k}{2(k-1)}}\frac{\Gamma(\frac{1}{2}+\frac{1}{k})}{\Gamma(\frac{1}{k})},
\end{equation}

\noindent
where $ m_{\phi}$ is the (scale dependent) inflaton mass, defined as the second derivative of the potential

\beq
m^2_\phi = V''(\phi_0)\,.
\eeq

\subsection{The gravitational wave spectrum}

Combining Eqs.~(\ref{Eq:inflatonpotential}) and (\ref{Eq:rhophi}), the frequency {\it at production} 
$f(a)=\frac{\omega(a)}{2 \pi}$ depends on the scale factor $a$ as

\beq
\omega(a) =\omega_{\rm end} \left(\frac{\ae}{a}\right)^\frac{3(k-2)}{k+2}
=\gamma_k~M_P~\left(\frac{\rhoe}{M_P^4}\right)^\frac{k-2}{2k}\left(\frac{\ae}{a}\right)^\frac{3(k-2)}{k+2},
\label{Eq:omegaphi}
\eeq
with
\beq
    \gamma_{k} = \sqrt{\frac{\pi}{2}}k\frac{\Gamma(\frac{1}{2}+\frac{1}{k})}{\Gamma(\frac{1}{k})}\lambda^{\frac{1}{k}}\,.
\eeq

This means that the GW frequency at production is constant for a quadratic potential ($k=2$), redshifts as the momentum
for $k=4$ ($\omega\propto \frac{1}{a}$), and redshifts {\it faster} than the expansion rate
for $k>4$. As a result, the present frequency of GWs produced at the scale factor $a$ by inflaton scattering can be computed as

\beq
f_0(a)=\frac{\omega(a)}{2 \pi}\frac{a}{a_0}=\frac{\gamma_k}{2 \pi}M_P \left(\frac{\rhoe}{M_P^4}\right)^\frac{k-2}{2k}
\frac{\ae^\frac{3k-6}{k+2}}{a_0}a^\frac{8-2k}{k+2} \,.
\label{Eq:f0inflaton}
\eeq

Before proceeding with the calculations, let us consider the expected shape of the GW spectrum resulting from inflaton scatterings. The qualitative features of the spectrum vary for different values of $k$.  For a quartic
potential $(k=4, w_\phi = \frac{1}{3})$, we find that the dominant inflaton mode redshifts as $\propto \frac{1}{a}$, leading to the 
scale factor dependence in Eq.~(\ref{Eq:f0inflaton}) effectively vanishing. This indicates that the GWs produced reinforce each other, resulting in a forest of monochromatic frequencies given by

\begin{equation}
n \, f_0 \sim n \times 0.27 \, \lambda^{\frac{1}{4}} \frac{\rhoe^{\frac{1}{4}}}{M_P} \frac{\ae}{a_0} \simeq 3 \, n \times 10^8 \, \text{Hz}.
\label{Eq:frequencyquartic}
\end{equation}

On the other hand, for a quadratic potential $(k=2, w_\phi = 0)$, the frequency at the time of production remains constant, given by $\omega = m_\phi$. This means that the present highest frequency  corresponds to the least redshifted modes, specifically those generated at the end of the reheating period, when the scale factor $a = \arh$.
At this stage, the inflaton density is at its most diluted state, resulting in a reduced GW production rate that is proportional to $\rho_\phi^2$. Consequently, we expect a spectrum that is tilted toward the infrared frequencies, contrasting with the spectrum of PBHs shown in Fig.~(\ref{Fig:gwfrompbh}).
The present peak frequency can be expressed as

\beq
\left.f_0^{\rm peak}\right|_{k=2}=m_\phi \frac{\ae}{a_0}\simeq 4.89\times 10^6\,\frac{m_\phi}{3\times 10^{13}}\left(\frac{\trh}{10^{10}~}\right)^\frac13\, {\rm Hz}\,,
\nonumber
\eeq
where we took $\rhoe=(4.8\times 10^{15})^{4}~{\rm GeV^4}$ \cite{Clery:2021bwz,Clery:2022wib,Choi:2024bdn}, and the units are in GeV if not specified. The highest frequency corresponds to gravitons produced just before the inflaton condensate is depleted and reheating completes, and reads

\beq
\left.f_0^{\rm max}\right|_{k=2} = m_\phi \, \frac{\arh}{a_0} \simeq 5.6\times 10^{13} \frac{10^{10}}{\trh}\,{\rm Hz}\,.
\eeq

For $k>4$ $(w_\phi>\frac13)$, we expect the largest rate of GW production to occur at the very end of inflation, corresponding to the highest {\it present} frequencies. This is because, as seen in Eq.~(\ref{Eq:f0inflaton}) for $k>4$, 
earlier production results in a larger $f_0$.

The spectrum $\frac{d\rho_{\rm GW}^\phi}{d \ln f_0}$ at the present time can be expressed as 

\bea
&&
\frac{d \rho_{\rm GW}^\phi(a_0)}{d f_0}=\frac{1}{a_0^4}\frac{d \left[a_0^4 \rho_{\rm GW}^\phi(a_0)\right]}{df_0}
=\frac{1}{a_0^4}\frac{d \left[a^4 \rho_{\rm GW}^\phi(a)\right]}{df_0}
\nonumber
\\
&&
=\frac{\sqrt{3}\,\rhoe^\frac32}{4M_P^3} \frac{k+2}{|k-4|}\left(\frac{2\pi f_0}{\omega_{\rm end}}\right)^\frac{3k-3}{k-4}\left(\frac{a_0}{\ae}\right)^\frac{9}{k-4} \Sigma^k \, ,
\nonumber
\eea
implying
\bea
&&
\Omega_{\rm GW}^\phi = \frac{1}{\rho^0_c}\frac{d\rho_{\rm GW}^\phi}{d\ln f_0}=
\frac{M_P^2}{4\sqrt{3}H_0^2}\frac{k+2}{|k-4|}\left(\frac{2 \pi}{\gamma_k}\right)^\frac{3k-3}{k-4}\frac{1}{\alpha^\frac{3k+6}{2k(k-4)}}
\nonumber
\\
&&
\times
\left(\frac{M_P}{\trh}\right)^\frac{6k+12}{k(k-4)}
\left(\frac{g_{RH}^\frac13}{g_0^\frac13}\frac{\trh}{T_0}\right)^\frac{9}{k-4} \Sigma^k
\left(\frac{f_0}{M_P}\right)^\frac{4k-7}{k-4}\,,
\label{Eq:omegagwphifinal}
\eea
where we combined Eqs.~(\ref{Eq:rhogwphi}) and (\ref{Eq:f0inflaton}) to obtain the relation between $a$ and $f_0$. We also 
used the fact that the GWs, in the case of inflaton scatterings, are produced at the very end of inflation, which implies $\rho_{\rm GW}^\phi\propto a^{-4}$ from $\ae$ until $a_0$. This result can be recovered by 
explicitly solving Eq.~(\ref{Eq:rhogwphi}), and has also been obtained in other contexts, such as leptogenesis, reheating, or gravitational dark matter 
production in Refs.~\cite{Clery:2021bwz,Clery:2022wib,Barman:2022qgt}. Physically, this scaling arises because the Planck-reduced production rate is too weak and redshifts too rapidly to counterbalance the Hubble expansion. Consequently, the gravitational waves are redshifted in the same way as a classical radiation field.

Note that for $k=4$, Eq.~(\ref{Eq:omegagwphifinal}) does not hold, because the redshift of the frequency of the source $\phi$, at a given time is identical to the redshift of the frequencies of the GWs produced before (both redshifts are proportional to $a^{-1}$). This results in a sum of monochromatic delta signals.

\subsection{Results}
\begin{figure*}[ht!]
    \centering
    \subfigure{\includegraphics[width=0.49\linewidth]{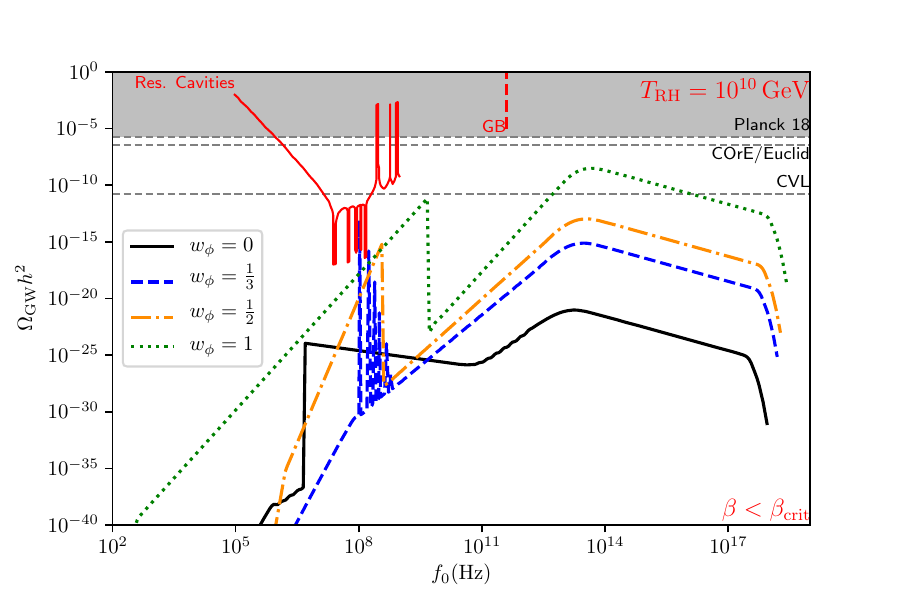}}
    \subfigure{\includegraphics[width=0.49\linewidth]{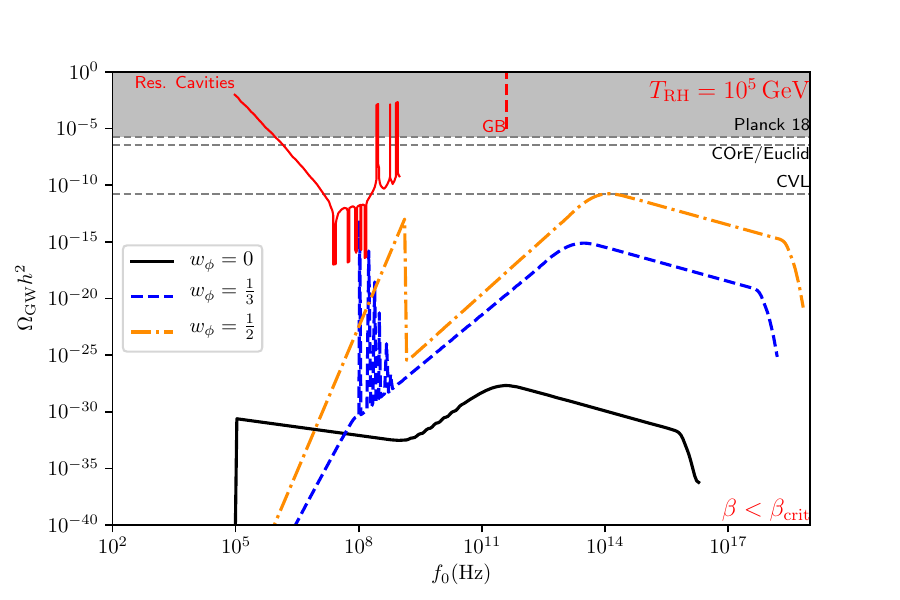}}
     \caption{Stochastic GW background for various equations of state considering inflaton scattering and a PBH population with  $\beta = 10^{-12}$ and $M_{\rm in} = 1$. We display the cases $\trh = 10^{10}$ GeV (left) and $\trh = 10^{5}$ GeV (right). Note that for $\trh= 10^{5}$ we did not plot the $w=1$ case due to PBH reheating effects, which we discuss in the following section. The gray shaded area is excluded by the Planck satellite, and the two additional gray lines indicate projected limits on $\Delta N_{\rm eff}$. We also 
    show the resonant cavities prospects, and experiments based on the GW electromagnetic wave 
conversion in a Gaussian beam (GB).
}
    \label{Fig:pbh+scattrh10}
\end{figure*}

We show in Fig.~(\ref{Fig:pbh+scattrh10}) the complete GW spectrum in the case where PBHs never dominate the energy density of the Universe. This spectrum, produced by the inflaton $\phi$ and the PBHs, is shown as a function of $f_0$ for the reheating temperatures $\trh = 10^{10}$ GeV (left) and $10^5$ GeV (right), setting $\beta = 10^{-12}$ and $\Min = 1$ g. Spectra for different values of $\beta$ result from a simple rescaling. We have verified that there are no PBH reheating effects for the benchmark scenarios shown in Fig.~(\ref{Fig:pbh+scattrh10}). Note that if PBHs dominate the energy budget of the Universe, their evaporation will drive the reheating. The $w=1$ case is excluded in Fig.~(\ref{Fig:pbh+scattrh10}) for this reason, and we will address it in the next section.

For 
the inflaton scattering 
contribution, we considered different values of $k$ ($k=2$, 4, 6 and $k\rightarrow \infty$, corresponding to $w_\phi=0, \frac13$, $\frac12$,
and $\sim 1$, respectively).\footnote{For $w_\phi=0$, $\gamma=0$, we then needed to fix it arbitrarily to 1 to allow for PBH formation.
} 
On top of the spectra, we display the various CMB constraints: the gray shaded region is excluded by Planck \cite{Planck2018Cosmologicalparam}, while the other two gray lines refer to future expected limits \cite{thecorecollaboration2011corecosmicoriginsexplorer,laureijs2011eucliddefinitionstudyreport,Ben_Dayan_2019} on the value of $\Delta N_{\rm eff}$. We also 
incorporate proposals for detection such as resonant cavities~\cite{Berlin_2022,berlin2023mago20electromagneticcavitiesmechanical,Herman_2023}, and experiments based on the GW electromagnetic wave 
conversion in a Gaussian beam (GB)~\cite{Li_2003}. 

Among the spectra presented, the one corresponding to the quartic potential ($w_\phi = \frac{1}{3}$) is easily distinguishable. The frequency at the moment of production redshifts as $\propto a^{-1}$,
resulting in each mode producing a delta peak at a frequency given by Eq.~ (\ref{Eq:frequencyquartic}). 
 
Additionally, we note the slope $\Omega_{\rm GW}^\phi \propto f_0^{\frac{4k-7}{k-4}}$ for the other cases, which aligns with our analytical result in Eq.~(\ref{Eq:omegagwphifinal}). For $k=2$, the lower cutoff corresponds to the most redshifted one, 
\beq
f_0^{\rm min}|_{k=2}=\frac{\omega_{\rm end}}{2\pi}\frac{\ae}{\arh}\frac{\arh}{a_0}
\simeq 4.9\times 10^6\left(\frac{\trh}{10^{10}{\rm GeV}}\right)^\frac13\,{\rm Hz}\,,
\eeq
where we have used Eq.~(\ref{Eq:omegaphi}) for $\omega_{\rm end}$. For $\trh=10^5 \, \rm{GeV}$, we observe that this result  aligns with our numerical findings in Figs.~(\ref{Fig:pbh+scattrh10}).

For $k>4$, the spectrum has an UV cut-off
{\rm stops} at higher frequencies 

\beq
f_0^{\rm max}|_{k>4}=\frac{\omega_{\rm end}}{2\pi}\frac{\ae}{\arh}\frac{\arh}{a_0}=\alpha^\frac{k+2}{6k}\frac{g_0^\frac13}{g_{\rm RH}^\frac13}\frac{T_0}{\rhoe^\frac{k+2}{6k}}\trh^\frac{4-k}{3k}\,,
\label{Eq:f0phimax}
\eeq
or, specifically for $k=6$ ($w_\phi=\frac{1}{2}$),

\beq
f_0^{\rm max}|_{k=6}\simeq3.7\times 10^8\left(\frac{\trh}{10^{10}~{\rm GeV}}\right)^{-\frac19}\,{\rm Hz}\,,
\eeq
and for large values of $k$ ($w_\phi=1$), we obtain
\beq
f_0^{\rm max}|_{k\gg 1}\simeq 4.7\times 10^{10}\left(\frac{\trh}{10^{10}~\rm GeV}\right)^{-\frac13}\,{\rm Hz}.
\eeq

\subsection{Peak gravitational wave amplitude from inflaton scattering}

Concerning the GW spectrum produced by the inflaton,
Eq.~(\ref{Eq:omegagwphifinal}) gives

\beq
\Omega_{\rm GW}^\phi\sim \trh^\frac{3k-12}{k(k-4)}\,f_0^\frac{4k-7}{k-4}\,,
\eeq
which results in $\trh^\frac32\,f_0^{-\frac12}$ for $k=2$ and $\trh^\frac12\,f_0^\frac{17}{2}$ for $k=6$.
It reaches its maximum at $f_0^{\rm min}$ for $k=2$, and $f_0^{\rm max}$
for $k>4$. 
Combining Eq.~(\ref{Eq:omegagwphifinal}) with Eq.~(\ref{Eq:f0phimax}), we see that, at the peak,

\beq
\Omega_{\rm GW}^\phi(f_0^{\rm peak})\sim \trh^\frac{16-4k}{3k}\,,
\label{Eq:omegagwphipeak}
\eeq
or

\bea
&&
\Omega_{\rm GW}^\phi(f_0^{\rm peak})|_{k=2}\simeq 1.2\times 10^{-24}\,\left(\frac{\trh}{10^{10}~{\rm GeV}}\right)^\frac43\, \rm \,,
\nonumber
\\
&&
\Omega_{\rm GW}^\phi(f_0^{\rm peak})|_{k=6}\simeq 6.4\times 10^{-16}\,\left(\frac{10^{10}~\rm GeV}{\trh}\right)^\frac49~\rm \,,
\nonumber
\\
&&
\Omega_{\rm GW}^\phi(f_0^{\rm peak})|_{k\gg 1}\simeq 7\times 10^{-12}\,\left(\frac{10^{10}~\rm GeV}{\trh}\right)^\frac43\, \rm \,,
\nonumber
\eea
where $f_0^{\rm peak}$= $f_0^{\rm min}$ for $k=2$, and $f_0^{\rm max}$ otherwise.
These analytical results reproduce what we observe in Figs.~(\ref{Fig:pbh+scattrh10}).

Now we compare the behavior of the peak GW amplitude with respect to $\trh$ generated by PBHs, Eq.~ (\ref{Eq:omegapeakpbh}), and the one generated by inflaton scatterings, Eq.~(\ref{Eq:omegagwphipeak}). Rewriting Eq.~(\ref{Eq:omegapeakpbh}), we get
\beq
\Omega_{\rm GW}^{\rm BH}(f_0^{\rm peak}) \sim \frac{\Min^\frac{4}{k}}{\trh^\frac{4k-16}{3k}}\,.
\eeq
We observe a ``miraculous" cancellation of the dependence on the reheating temperature in the ratio at the peak,

\beq
\label{eq:ratio}
\frac{\Omega_{\rm GW}^{\rm BH}}{\Omega_{\rm GW}^\phi}\sim  \beta \,\Min^\frac{4}{k}\,.
\eeq
This gives

\bea
&&
\left.\frac{\Omega_{\rm GW}^{\rm BH}}{\Omega_{\rm GW}^\phi}\right|_{k=2}\simeq 3.6\times10^4\,
\left(\frac{\beta}{10^{-12}}\right)\,\left(\frac{\Min}{1~\rm g}\right)^2\,,
\nonumber
\\
&&
\left.\frac{\Omega_{\rm GW}^{\rm BH}}{\Omega_{\rm GW}^\phi}\right|_{k=6}\simeq 1760\,
\left(\frac{\beta}{10^{-12}}\right)\,\left(\frac{\Min}{1~\rm g}\right)^\frac23\,,
\nonumber
\\
&&
\left.\frac{\Omega_{\rm GW}^{\rm BH}}{\Omega_{\rm GW}^\phi}\right|_{k\gg 1}\simeq 4787\,
\left(\frac{\beta}{10^{-12}}\right)\,,
\eea
which is also what we observe in Fig.~(\ref{Fig:pbh+scattrh10}). It is interesting to remark that, for an equation of state $w=1$ ($k\gg 1$), the ratio is uniquely fixed by $\beta$. Eq.~(\ref{eq:ratio}) is one of the main results of our comparative study.
In fact, it does not only shows that the presence of GWs induced by PBH 
evaporation is unavoidably connected to the presence of GWs due to the background itself, but also that the ratio of their peak amplitudes is a clear signature of the nature of the background, being different for different values of $w_\phi$.

It is complementary to study the GWs produced indirectly, by the
radiative bremsstrahlung of gravitons from the products of inflaton decay, as it was done in \cite{Barman:2023ymn,Bernal:2023wus,Barman:2023rpg}, and more recently in \cite{Choi:2024acs}, where the authors considered the bremsstrahlung of gravitons from the heavy products of PBH decay. We do not include these contributions in our analysis for several reasons. First, we are interested in the {\it minimal} and {\it unavoidable} amount of GWs produced by PBHs plunged into a bath populated by the inflaton and radiation.  Second, the radiative bremsstrahlung depends strongly on the type of particles produced (scalars, fermions or vectors) and their masses, with the heavier ones contributing much more to the radiative bremsstrahlung than the lighter ones. Finally, this second order effect also depends strongly on the process that generates the bremsstrahlung (decay, scattering...), and on the coupling of the inflaton to the particles of the SM. In our study, we have restricted ourselves to pure gravitational couplings to keep a conservative approach.

\section{Primordial black hole domination}
\label{sec:pbhdomination}

\subsection{Primordial black hole reheating}

At this stage, we have considered the case of PBHs immersed in a Universe dominated 
by the inflaton condensate. However, it was shown in \cite{RiajulHaque:2023cqe,Haque:2023awl,Barman:2024slw} that there exist the possibility for the PBHs to dominate the energy budget of the Universe before the end of reheating. Indeed, 
this can happen in two cases:

\begin{itemize}
\item{PBH domination due to rapid redshifting of inflaton energy density: This situation occurs when the inflaton energy density, $\rho_\phi\propto a^{-\frac{6k}{k+2}}$, redshifts faster than the PBH energy density $\rho_{\rm BH}\propto a^{-3}$, and it becomes subdominant {\it before} the evaporation of the PBHs. This requires $k>2$ and a sufficiently large value of $\beta$.}
\item{PBH domination due to inflaton decay before PBH evaporation: This scenario occurs when the inflaton decays before the PBHs do, and the PBHs dominate the 
energy density of the radiation ($\propto a^{-4}$) {\it before} their evaporation.
}
\end{itemize}

\noindent
Once the PBHs are formed, their subsequent evolution depends on the nature of the background field. Here, we focus on the first scenario, where the PBH energy density eventually dominates over the energy density of the inflaton condensate.
 As it was shown in \cite{RiajulHaque:2023cqe,Haque:2023awl, Domenech:2024wao}, black holes formed in a Universe filled by the inflaton condensate can dominate the energy density of the universe provided that $\beta$ is larger than some critical value, $\beta_{\rm crit}$, given by

\beq
\rho_{\rm BH}(\aev)>\rho_\phi(\aev)
~~\Rightarrow ~~\beta>\beta_{\rm crit}=\left(\frac{H(\aev)}{H(\ain)}\right)^\frac{2w_\phi}{3(1+w_\phi)}\,,
\eeq
where we considered a Universe dominated by the inflaton before PBH domination, or
$H \propto a^\frac{-3(1+w_\phi)}{2}$. Taking 
$H(\aev)=\frac{2}{3(1+w_\phi)\tev}$, 
with $\tev$ given by Eq.~(\ref{Eq:tev}), and $H(\ain)=\Hin$
from Eq.~(\ref{Eq:min}), we obtain

\begin{eqnarray}
\label{Eq:betacrit} 
\beta_{\rm crit} & = & \left(\frac{\epsilon}{(1+w_\phi)2 \pi \gamma}\right)^{\frac{2w_\phi}{1+w_\phi}} \left(\frac{M_P}{\Min}\right)^{\frac{4w_\phi}{1+w_\phi}} \,.
\end{eqnarray}

\begin{figure}[H]
    \centering
    \includegraphics[width=1\linewidth]{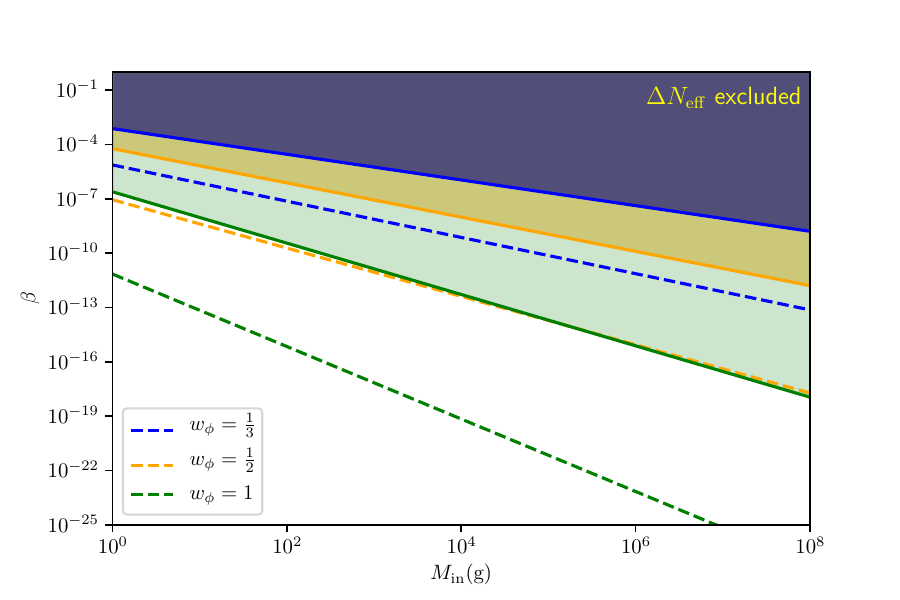}
    \caption{$\beta$ as a function of $M_{\rm in}$ for different values of $w_\phi=\frac13$ (blue), 
    $\frac12$ (orange) and 1 (green). 
    The dashed lines represent $\beta_{\rm crit}$, i.e.~the values above which the reheating is dominated by the population of PBHs over $\rho_\phi$. The colored shaded areas above the solid lines are excluded by the requirement $\Delta N_{\rm eff} \leq 0.3$, limit given by Eq.~(\ref{upper-bound-beta-updated})
    .}
    \label{Fig:betacrit}
\end{figure}

We show in Fig.~(\ref{Fig:betacrit}) (dashed lines), the value of $\beta_{\rm crit}$ 
for $w_\phi = \frac13, \,\frac12$ and 1.
%This corresponds to 
The upper bound on $\beta$ is derived from 
the limit on $\Delta N_{\rm eff}$ ~\cite{Inomata:2020lmk,Hooper:2020evu,Papanikolaou:2020qtd,Domenech:2020ssp,domènech2024formationevaporationinducedgravitational}. Here we used the bound recently derived in Ref.~~\cite{Domenech:2024wao}, where the Universe undergoes PBH formation during a $w$-dominated era, prior to PBH domination 
\begin{multline} \label{upper-bound-beta-updated}
\beta < (2.3\times10^{-32})^{\frac{3w}{4(1+w_\phi)}}C(w)^{-\frac{3w_\phi}{1+w_\phi}}\left(\frac{g_{BH}}{108}\right)^{\frac{17w_\phi}{12(1+w_\phi)}} 
\\
\left(\frac{\gamma}{0.2}\right)^{-\frac{2w_\phi}{1+w_\phi}}  \left(\frac{M_{\rm in}}{10^{4}\, {\rm g}} \right)^{-\frac{17w_\phi}{6(1+w_\phi)}} \, ,
\end{multline}

where $C(w)$ is defined as
\beq
C(w) = \frac{9}{20}\alpha_{\rm fit}^{-\frac{1}{3w}}\left(3+\frac{1-3w}{1+3w}\right)^{-\frac{1}{3w}} \, ,
\eeq
and $\alpha_{\rm fit} \simeq 0.135$ is a constant computed numerically.
As we can see in Fig.~(\ref{Fig:betacrit}), a stiffer equation of state parameter, $w_\phi$, leads to a significantly stronger upper bound on $\beta$.
Indeed, for larger values of $w_\phi$, or larger pressure, 
the GW amplitude is amplified by the prolonged phase of PBH domination~\cite{Domenech:2024wao}. As $w_\phi$ increases, the allowed range for $\beta$ shifts towards lower values, and the allowed parameter space narrows as $w_\phi \to 1$. In contrast, lower values of $w_\phi$ relax this bound. 
%here 
Additionally, note also that larger initial PBH masses, correspond to lower allowed values of $\beta$. This is due to the fact that a late decay of PBHs (corresponding to larger $\Min$)
induces less dilution on the GW spectrum, resulting in a more constrained parameter space.

In the case where $\beta>\beta_{\rm{crit}}$, the reheating is driven by the sudden evaporation of the PBHs, and $\trh$ reads~\cite{RiajulHaque:2023cqe}

\beq
H_{\rm RH} = \Gamma_{\rm BH}=\frac{\epsilon M_P^4}{\Min^3}~~\Rightarrow ~~\trh=
\left(\frac{3\,\epsilon^2}{\alpha}\right)^{\frac{1}{4}} \left(\frac{M_P}{M_{\rm in}}\right)^{\frac{3}{2}} M_P\,.
\label{Eq:trhbetac}
\eeq
Note that different definitions of the reheating temperature can vary by factors of order unity. For example, if we define $\trh$ by the condition $H_{\rm RH}=H(\aev)$, we obtain a reheating temperature $\sqrt{2}$ times larger. This uncertainty in the definition also arises from the fact that the reheating medium cannot be defined by a single component, but by the sum of two components.

Another interesting point is that $\trh$ does not depend on $\beta$. This is analogous to the case where the reheating temperature from inflaton decay is independent from $\rhoe$. Only the ``time'' of decay, defined by $\Gamma_\phi$ or $\Gamma_{\rm BH}$, determines the size of the universe at the radiation-dominated epoch, and thus the radiation density $\rho_{\rm RH}$. In other words, once the PBHs dominate the energy budget of the Universe ($\beta>\beta_{\rm crit}$), the reheating temperature $\trh$ is no longer an 
independent parameter and is uniquely determined once we know $\Min$. 
As a result, the redshift of the GWs from the end of reheating to the present time, $\propto T_0^3/\trh^3$, 
is uniquely determined by $\Min$, and is proportional to 
$\Min^{-\frac{9}{2}}$.

Finally, defining $\abh$ as the scale factor when $\rho_\phi(\abh)=\rho_{\rm BH}(\abh)$,
it is easy to show that

\bea
\frac{a_{\rm in}}{a_{\rm BH}}=\beta^{\frac{1}{3 w_\phi}}\, ,
\label{Eq:ainabh} 
\eea
%%%%
\bea \label{Eq:aevabh}
\frac{a_{\rm RH}}{a_{\rm BH}} = \frac{M_{\rm in}^2\,\rho_{\rm in}^{\frac{1}{3}}}{(2 \sqrt{3}\epsilon M_P^5)^\frac{2}{3}} \beta^{\frac{1+w_\phi}{3 w_\phi}}\,,
\eea

leading to
%%%%%%
\bea \label{Eq:arhain} 
\frac{a_{\rm in}}{a_{\rm RH}} = \frac{1}{\beta^\frac{1}{3}}\left(\frac{\epsilon}{2\,\pi\,\gamma}\right)^{\frac{2}{3}}\left(\frac{M_P}{M_{\rm in}}\right)^{\frac{4}{3}}\,.
\eea

\subsection{The spectrum from PBH evaporation}

\begin{figure}[H]
    \centering
    \includegraphics[width=\linewidth]{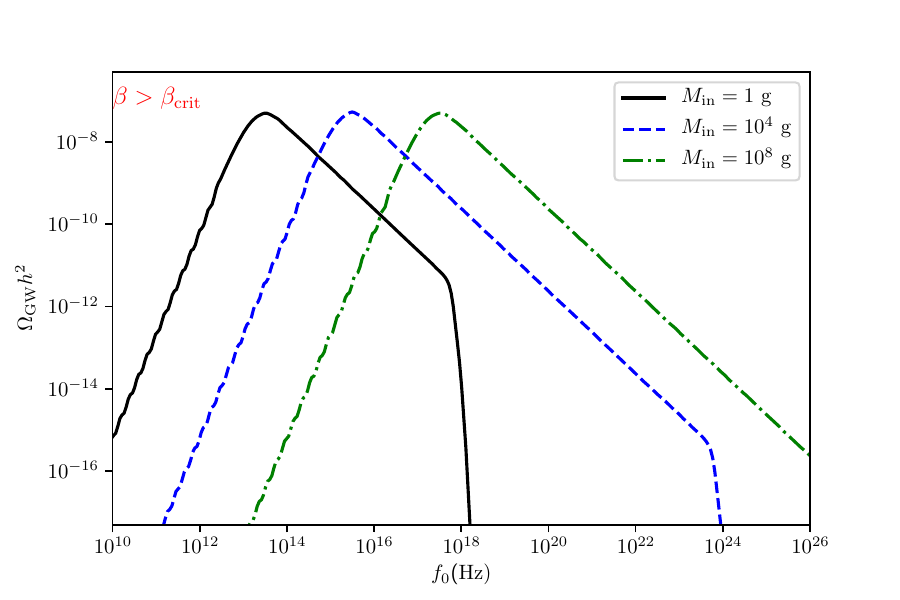}
    \caption{Stochastic GW background from PBH evaporation when they come to dominate the energy density of the Universe ($\beta>\beta_{\rm{crit}}$) for different masses, 
    $\Min=1$ (black), $10^4$ (blue) and $10^8$ g (green).}
    \label{Fig:pbhbetac}
\end{figure}

We show in Fig.~(\ref{Fig:pbhbetac}) the GW spectrum
directly emitted through PBH decay, for $\Min=1$, $10^4$ and $10^8$ g, and $\beta>\beta_c$. Our numerical results obtained with \texttt{BlackHawk} agree with \cite{Ireland:2023avg,Choi:2024acs}. 
We recover the value of the peak frequency 

\beq
k_0^{\rm peak}\simeq 2.8\times T_{\rm BH}\frac{\arh}{a_0}~~\Rightarrow f_0^{\rm peak}\simeq3\times 10^{13}\sqrt{\frac{\Min}{1~\rm g}}~\rm{Hz}\,, 
\label{Eq:k0peakbetac}
\eeq
where we used $\aev=\arh$ when $\beta>\beta_c$, $T_{\rm BH}=\frac{M_P^2}{\Min}$, and 
Eq.~(\ref{Eq:trhbetac}) for $\trh$. This value matches with our numerical results obtained in  Fig.~(\ref{Fig:pbhbetac}).
Similarly to the case where $\beta<\beta_c$, we recover the 
behavior $\Omega_{\rm GW}^{\rm BH} \propto f_0^3$ for $f_0<f_0^{\rm peak}$.

An interesting feature of the spectrum is that it is independent of 
$w_\phi$, $\beta$ and $\Min$ at the peak. This can easily
be undestood by adapting Eq.~(\ref{Eq:spectrumaev}) to the case $\beta>\beta_{\rm{crit}}$, integrating from $\abh$ (the onset of PBH domination)
to $\aev$. In this case, $\aev$ can be identified
with $\arh$, and

\beq
\left.\frac{d \rho}{d \ln k_0}\right|_{\beta>\beta_c}=
\frac{27 \sqrt{3}}{64 \pi^3}\frac{k_0^\frac52\sqrt{\rho_{\rm RH}}}{\sqrt{\Min}}\left(\frac{\aev}{a_0}\right)^\frac32\times I_0^{\beta_c}
\,,
\label{Eq:rhogwbhbetac}
\eeq
with

\beq
I_0^{\beta_c}=\int_{\frac{k_0 \Min}{M_P^2}\frac{a_0}{\aev}}^{\frac{k_0 \Min}{M_P^2}\frac{a_0}{\abh}}\frac{\sqrt{Y}}{e^Y-1}dY\,.
\eeq
It is easy to see that $I_0$ does not depend on $\beta$ or $\Min$. In fact, only the lower bound of the integral gives the main 
contribution to the decay (most of the spectrum is obtained at the end of the evaporation process), and the peak is reached for $k(\aev)=k_0\frac{a_0}{\aev}\sim T_{\rm BH}\sim\frac{M_P^2}{\Min}$, which makes the lower limit of the integral independent of $\beta$ and $\Min$. Using Eq.~(\ref{Eq:trhbetac}) for $\trh$,
we get

\bea
&&
\left.\frac{d \rho_{\rm GW}}{d \ln k_0}\right|_{\beta>\beta_c}=
\frac{27}{64 \pi^3}\sqrt{\frac{3 \alpha g_0}{g_{\rm RH}}}
\left(\frac{3 \epsilon^2}{\alpha}\right)^\frac18\left(\frac{M_P}{\Min}
\right)^\frac54k_0^\frac52T_0^\frac32 I_0^{\beta_c}
\,.
\nonumber
\eea
Evaluating the expression above at $k_0=k_0^{\rm peak}$, 
we obtain

\beq
\left.\frac{d \Omega_{\rm GW}^{\rm BH}}{d \ln k_0}\right|_{\beta>\beta_c}^{\rm peak}=
\frac{27}{64 \pi^3}\frac{\alpha}{\epsilon}\left(\frac{g_0}{g_{\rm RH}}\right)^\frac43 \frac{T_0^4}{3 M_P^2H_0^2} \times I_0^{\beta_c}\simeq 8.5\times 10^{-8}\,.
\nonumber
\eeq
This clearly shows that the peak amplitude remains independent of $\beta$ and $\Min$, as we observe in Fig.~(\ref{Fig:pbhbetac}). 
Our analytical approximation is in good agreement with our numerical results. This is expected because once PBHs dominate the universe, the history from the end of inflation until $a_{\text{BH}}$ is effectively ``erased'' and, hence, $\omega_\phi$ or $\beta$ should not affect the physics from $a_{\text{BH}}$ until the evaporation time.

\subsection{Contribution from density fluctuations}

When $\beta>\beta_{\rm{crit}}$, one needs to take into account a third source of gravitational waves, indirectly generated through the presence of PBHs. It is known that
the initial inhomogeneous distribution of PBHs causes isocurvature fluctuations, which, in turn, generate  curvature perturbations as the Universe evolves \cite{Papanikolaou:2020qtd}. When PBHs dominate the Universe energy density, perturbations can grow faster due to the pressureless environment. Then, the non-linear nature of gravity converts these scalar fluctuations into tensor
perturbations, leading to GW production. More precisely, 
the near-instantaneous PBH evaporation, modeled as sudden reheating, transforms large density fluctuations into radiation, producing significant pressure waves and a quadrupole moment, which leads to GW production~\cite{Domenech_2021}.

When PBHs form, only the largest primordial fluctuations from the distribution tail exceed the required threshold to trigger PBH formation. Therefore, PBHs can be assumed to be randomly distributed throughout the Universe, following a Poisson spectrum for the density fluctuations. The real-space two-point function for the density contrast \cite{Papanikolaou:2020qtd} can be written as

\bea
\biggl<\frac{\delta \rho_{\rm BH}(\bf x)}{\rho_{\rm BH}} \frac{\delta \rho_{\rm BH}(\bf x^\prime)}{\rho_{\rm BH}} \biggr> = \frac{4 \pi}{3} \left(\frac{d}{a}\right)^3 \delta({\bf x} - {\bf x'})\, ,
\label{Eq:deltarhox}
\eea
where $d$ denotes the mean spatial separation between PBHs. 
Demanding $\frac{4 \pi}{3}\pi \din^3\times \nin =1$ at the time of
formation, one obtains the distance

\bea
&&
\din= \left(\frac{\gamma}{\beta}\right)^{\frac{1}{3}} H_{\rm in}^{-1}\,,
\label{Eq:initial_diki}
\eea
with $\din=d(\ain)$. Following this, we can compute the current peak frequency $f_0^{\rm peak}$ of the GW spectrum from  density fluctuations

\beq
f_0^{\rm peak}=f_{\rm in} \times \left(\frac{\ain}{\aev}\right)\left(\frac{\aev}{a_0}\right)\,,
\eeq
where $f_{\rm in}=\frac{1}{2 \pi}\din^{-1}$, $\frac{\ain}{\aev}$ is 
given by Eq.~(\ref{Eq:arhain}) with $\aev \equiv \arh$, and $\frac{\aev}{a_0}=\left(\frac{g_0}{g_{\rm RH}}\right)^\frac13\frac{T_0}{\trh}$. Next, using the expression of  
$\trh$ given in Eq.~(\ref{Eq:trhbetac}), we obtain

\bea
f_0^{\rm peak}=&&\frac{1}{2\pi}
\times
4 \pi \gamma \frac{M_P^2}{\Min}
\left(\frac{\beta}{\gamma}\right)^\frac13
\times
\frac{1}{\beta^\frac{1}{3}}\left(\frac{\epsilon}{2\,\pi\,\gamma}\right)^{\frac{2}{3}}\left(\frac{M_P}{M_{\rm in}}\right)^{\frac{4}{3}}
\nonumber
\\
&&
\times
\left(\frac{g_0}{g_{RH}}\right)^\frac13
\left(\frac{\alpha}{3\,\epsilon^2}\right)^{\frac{1}{4}} \left(\frac{M_{\rm in}}{M_P}\right)^{\frac{3}{2}} \frac{T_0}{M_P}
\nonumber
\\
&&
=\frac{4 \pi~\epsilon^\frac16}{(2 \pi)^\frac23}
\left(\frac{g_0}{g_{RH}}\right)^\frac13
\left(\frac{\alpha}{3}\right)^\frac14
\left(\frac{M_P}{\Min}\right)^\frac56~T_0
\nonumber
\\
&&
\simeq
2.5\times 10^3 \left(\frac{10^4}{\Min}\right)^\frac56~\rm Hz
\label{Eq:f0peakdeltarho}
\,,
\eea
where we used Eq.~(\ref{Eq:min}) to express $\Hin$ as function of $\Min$.
For $10^4~ {\rm g} \lesssim \Min \lesssim 10^8~ {\rm g}$, the range of frequencies around $f_0^{\rm peak}$ is within the observable
region of LIGO/VIRGO/KAGRA \cite{Abbott_2021} and future detectors such as ET~\cite{Hild_2011}.\footnote{The slight difference in the normalization of $f_0^{\rm peak}$ compared to \cite{Domenech:2020ssp} is due to the different gray-body factor we used.}

\begin{figure}[H]
\includegraphics[width=\linewidth]{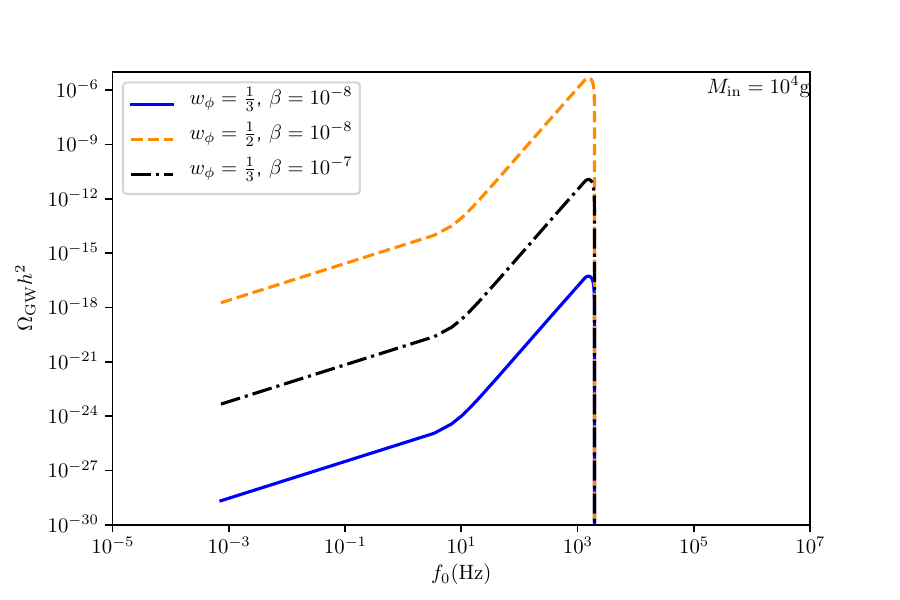}
\caption{Gravitational wave spectrum induced by PBH density fluctuations. The three curves are computed for $M_{\rm in} = 10^4 \, \text{g}$ and an initial PBH fraction of $\beta=10^{-8}$ (blue and orange), 
and $\beta=10^{-7}$ (black). We chose $w_\phi=\frac{1}{3}$ (blue and black), and  $w_\phi=\frac{1}{2}$ (orange). 
}
\label{fig:ind_GWs}
\end{figure}

\begin{figure}[H]
    \centering
    \includegraphics[width=\linewidth]{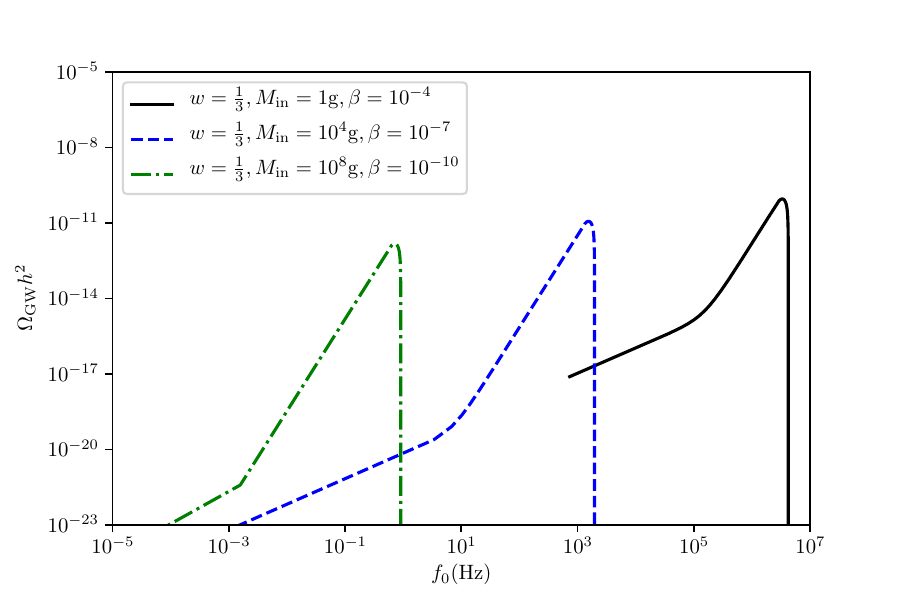}
    \caption{Gravitational wave spectrum induced by PBH density fluctuations for different values of $\Min$ and $k=4$ ($w=1/3$).}
    \label{fig:isocurvature_spectra}
\end{figure}

Note that Eq.~(\ref{Eq:f0peakdeltarho}) does not depend on either $\beta$ or $\gamma$.  Indeed, as discussed in the previous section, once the PBH population dominates the energy budget, the prior history is effectively erased, and the 
consequences depend uniquely on the PBH lifetime $\Gamma_{\rm ev}^{-1}$,
and not on their initial density. We also note that
the peak frequency {\it decreases} $\propto\Min^{-\frac56}$ with $\Min$, which is the 
case for PBH evaporation, where $f_0^{\rm peak}$ increases with $\Min$.
To illustrate this, we show in Figs.~(\ref{fig:ind_GWs}) and (\ref{fig:isocurvature_spectra})
the GW 
spectra induced by the density perturbations for different values of $\Min$, which effectively peak 
at the value of the frequency $f_0^{\rm peak}$ obtained in Eq.~(\ref{Eq:f0peakdeltarho}).

The GW spectrum is expected to be truncated for frequencies below the size of the horizon at the time of evaporation, $k_{\rm RH}=H_{\rm ev}=H_{\rm RH}= \Gamma_{BH}$. Indeed, such a long wavelength should not be (causally) affected by any events taking place within a volume of horizon size.
From Eq.~(\ref{Eq:trhbetac}), we can deduce the current cut-off frequency 

\beq
f_{\rm RH}=\frac{H_{\rm RH}}{2 \pi}\left(\frac{g_0}{g_{\rm RH}}\right)^\frac13\frac{T_0}{\trh}
\simeq 7\times 10^{-4}\left(\frac{10^4~\rm g}{\Min}\right)^\frac32~{\rm Hz}\,,
\label{Eq:frh}
\eeq
a result which aligns with what we observe in Fig.~(\ref{fig:isocurvature_spectra}).
In order to obtain the full shape of the spectrum, we should transform
the isocurvature density fluctuations of the PBHs into gravitational
curvature fluctuations, which in turn generate the
the second order GWs. This is the procedure we will describe in more detail below.

As mentioned above, due to the discrete distribution of PBHs, the fluctuations are constrained by the mean separation between PBHs. We then set the UV cut-off for the comoving wavenumber as
\bea
k_{\rm UV} = \frac{a_{\rm in} }{\din} = \left(\frac{\beta}{\gamma}\right)^{\frac{1}{3}} \kin \, ,
\label{Eq:UVcutoff}
\eea
with $\kin = a_{\rm in} H_{\rm in}$ (see Eq.~(\ref{Eq:initial_diki})).
Now, from the initial isocurvature perturbation $S_i \approx \delta \rho_{\rm BH} / \rho_{\rm BH}$, 
we can calculate the initial power spectrum 
$P_{{\rm BH},i}(k)$
by a Fourier expansion of the density contrast (\ref{Eq:deltarhox})~\cite{Papanikolaou:2020qtd,Domenech:2020ssp}

\bea
\biggl<\frac{\delta \rho_{\rm BH}(\bf x)}{\rho_{\rm BH}} \frac{\delta \rho_{\rm BH}(\bf x^\prime)}{\rho_{\rm BH}} \biggr> = \int\frac{d^3 k}{(2 \pi)^3} P_{{\rm BH},i}(k)e^{{\bf k}.({\bf x}-{\bf x'})} \, ,
\label{Eq:deltarhok}
\eea
with
\bea
P_{{\rm BH},i}(k) = \frac{4 \pi}{3} \left(\frac{d}{a} \right)^3\,.
\eea
We then define the {\it reduced} power spectrum

\beq
\biggl<\frac{\delta \rho_{\rm BH}(k)}{\rho_{\rm BH}} \frac{\delta \rho_{\rm BH}(k^\prime)}{\rho_{\rm BH}} \biggr>=\frac{2 \pi^2}{k^3}{\cal P}_{{\rm BH},i}(k)\delta(k+k')\,,
\eeq
or
\bea
\mathcal{P}_{{\rm BH},i}(k) = \frac{k^3}{2 \pi^2}P_{{\rm BH},i}=
\frac{2}{3\pi} \left(\frac{k}{k_{\rm UV}} \right)^3\,.
\label{Eq:isocurvature}
\eea

The initial isocurvature fluctuations $S_i$, powered by ${\cal P}_{{\rm BH},i}(k)$, eventually transform into {\it curvature} perturbations due to the coupled evolution for $S_i$ and the gravitational potential $\Phi$,
which is defined in the Newton gauge for the metric as

\beq
ds^2=(1-2\Phi)dt^2-a^2(t)\left[\delta_{ij}(1+2\Phi)+\frac{2~ h_{ij}}{M_P}\right]dx^i dx^j\,,
\eeq
where $h_{ij}$ is the transverse-traceless component of the spatial metric.
 This, in turn, generates a curvature (adiabatic) power spectrum for $\Phi$, $\mathcal{P}^2_{\Phi}(k)$, which then sources the production of second-order GWs, 
 associated to the power spectrum ${\cal P}_h(k)$. 
 The GW spectral density, $\Omega_{\rm GW}(k)$, can be derived from the tensor power spectrum, $\mathcal{P}_h(k)$. Here we will use the parametric formulas derived in~\cite{Domenech:2024wao} to compute $\Omega_{\rm GW}(k)$. The analysis of the evolution of fluctuations within a general fluid with an equation of state $w_\phi$ is beyond the scope of this work, and we refer the reader to \cite{Domenech:2024wao} for further details.

Near the resonant peak the GW spectral density can be estimated as
\begin{equation}
    \label{eq:OmegaGWres}
    \Omega_{\rm GW}^{\rm peak}(k) 
    \approx
    \Omega_{\rm GW, res}^{\rm peak}
    \left(\frac{k}{k_{\rm UV}}\right)^{11/3}
    \Theta(k_{\rm UV}-k),
\end{equation}
with $\Theta$ the heaviside function which acts as a cutoff of the GW 
spectrum for high frequencies (corresponding to the present frequency $f_0^{\rm peak}$ discussed above), 
and where the amplitude near the peak is expressed as
\footnote{This analytical parametrization underestimates the peak amplitude by a factor of 2 \cite{Domenech:2024wao} this extra factor is taken into account in our plots}

\begin{align}  
\Omega_{\rm GW, res}^{\rm peak} &= C^4(w)\frac{c_s^{7/3}(c_s^2-1)^2}{576 \times 6^{1/3}\pi}
\left(\frac{k_{\rm BH}}{k_{\rm UV}}\right)^8
\left(\frac{k_{\rm UV}}{k_{\rm RH}}\right)^{17/3}
\label{eq:OmegaGWresPeak}
\\
&\approx 9.58 \times 10^{30} C^4(w) \beta^{\frac{4(1+w)} {3w}}\left(\frac{g_H}{108}\right)^{-17/9}
\nonumber
\\
&\hspace{4cm}
\times\left(\frac{\gamma}{0.2}\right)^{8/3}
\left(\frac{M_{\rm in}}{10^4\text{g}}\right)^{34/9} \,,
\nonumber
\end{align}
with
\beq
c_s=\frac{4}{9}\frac{\rho_R}{\rho_{\rm BH}+\frac43\rho_R}\,.
\eeq

Here $k_{\rm BH}$ is the horizon size when the PBHs begin to dominate
the energy budget of the Universe, with

\beq
k_{\rm BH}= \sqrt{2}\beta^{\frac{1+3w}{6w}} k_{\rm in}\,,
\eeq
corresponding to the present frequency

\beq
f_{\rm BH}=\sqrt{2}\beta^{\frac{1+3w}{6w}} f^{\rm peak}_{0} \,.
\label{Eq:f0bh}
\eeq

We derive the GW signal $\Omega_{\rm GW}(k)$ by interpolating between the approximation $\Omega_{\rm GW}^{\rm peak}(k)$ at the resonant peak and the dominant contribution to the signal in the infrared tail~\cite{Domenech:2024wao}
\begin{align}
    \Omega_{\rm GW, IR}(k) = C^4(w) \frac{c_s^4 }{120 \pi ^2}\left(\frac{2}{3}\right)^{1/3} 
    &\left(\frac{k_{\rm BH}}{k_{\rm UV}}\right)^8  \\
    &\times \left(\frac{k_{\rm UV}}{k_{\rm RH}}\right)^{14/3}
    \left(\frac{k}{k_{\rm UV}}\right) \nonumber \label{eq:OmegaGWIR_Num}
\end{align}

\begin{align}
\approx 9.03 \times 10^{24} C^4(w) \beta^{\frac{4(1+w)}{3w}} &\left(\frac{g_H}{108}\right)^{-14/9} \left(\frac{\gamma}{0.2}\right)^{8/3} \\ &\times \left(\frac{M_{\rm in}}{10^4 \text{ g}}\right)^{28/9} \left(\frac{k}{k_{\rm UV}}\right)\,.
\nonumber
\end{align}

In Fig.~(\ref{fig:ind_GWs}) we show the GW spectra for an initial PBH mass of $M_{\rm in}=10^4 \, \text{g}$ and an initial PBH fraction of $\beta= 10^{-8}$ and $10^{-7}$ for two different values of $w_\phi=\frac13$ and $\frac12$. The position of the peak, $f_0^{\rm peak}$, does not depend on $\beta$ because 
the $\beta$ dependence of the initial comoving wave number cancels out once redshift is considered, see Eq.~(\ref{Eq:f0peakdeltarho}). 
However, the amplitude of the wave is indeed very sensitive to the value of $w_\phi$ since it enters in the exponent of $\beta$, see  Eq.~(\ref{eq:OmegaGWresPeak}), and typically $\beta \ll 1$. 
The larger the value of $\beta$, the smaller is the value of $d_{i}$ (more PBH are formed), 
see Eq.~(\ref{Eq:initial_diki}), and the longer is the PBH domination phase, driving the expansion of the Universe with a smaller pressure.

Conversely, larger values of $w_\phi$ result in a higher $k_{\rm BH}$ due to the overdilution of $\rho_\phi$. This leads to an increased GW density. The suppression effect $\Phi \propto a^{-2}$ between $a_{\rm UV}$ and 
$a_{\rm BH}$ is diminished because $a_{\rm BH}$ approaches $a_{\rm UV}$. Another way to understand this is by examining 
the form of the potential $\Phi$, which can be obtained after solving 
Einstein's equation for the gravitational potential \cite{Papanikolaou:2020qtd}, 

\beq
\Phi\simeq -\left(5+\frac49\frac{k^2}{k_{\rm BH}^2}\right)^{-1}\frac{\delta \rho_{\rm BH}}{\rho_{\rm BH}}\,,
\eeq
which means

\beq
{\cal P}_\Phi(k)=\frac{2}{3 \pi}\left(\frac{k}{k_{\rm UV}}\right)^3
\left(5+\frac49\frac{k^2}{k_{\rm BH}^2}\right)^{-2}\,.
\label{Eq:pphi}
\eeq
We can clearly see that for $k=k_{\rm UV}\gg k_{\rm BH}$,
${\cal P}_\Phi(k_{\rm UV})\propto \left(\frac{k_{\rm BH}}{k_{\rm UV}}\right)^4$, which yields a greater density of GW when $k_{\rm BH}$ goes to $k_{\rm UV}$, corresponding to larger values of $w_\phi$.

From Eq.~(\ref{Eq:pphi}), we distinguish two different regimes. Whereas we have
${\cal P}_\Phi \propto k^3$ for small frequencies, (corresponding to 
$k\ll k_{\rm BH}$), we obtain ${\cal P}_\Phi\propto \frac{1}{k}$ for $k\gg k_{\rm BH}$. We can clearly see the shift in the slope of the spectrum in Fig.~(\ref{fig:ind_GWs}) at the breaking point \cite{Domenech:2020ssp}

\beq
f_{\rm br}\simeq 10~{\rm Hz} \left(\frac{10~{\rm g}}{\Min}\right)^\frac{13}{12}\,.
\label{Eq:fbr}
\eeq

The dependence on the mass $\Min$ is well illustrated in Fig.~(\ref{fig:isocurvature_spectra}). We observe that the features of the spectrum shift to lower frequencies for larger masses, according to Eq.~(\ref{Eq:f0peakdeltarho}) ($\propto \Min^{-\frac56}$), due to the increased inter-PBH distance $\din$ at formation time, Eq.~(\ref{Eq:initial_diki}). 
In summary, the main features of the GW spectrum are its onset at $f_{\rm br}$, Eq.~(\ref{Eq:frh}), the change in slope at $f_{\rm BH}$, Eq.~(\ref{Eq:fbr}), and its peak at $f_0^{\rm peak}$, Eq.~(\ref{Eq:f0peakdeltarho}).

\subsection{Summary}

\begin{figure*}
    \centering
    \subfigure{\includegraphics[width=\textwidth,height = 10cm]{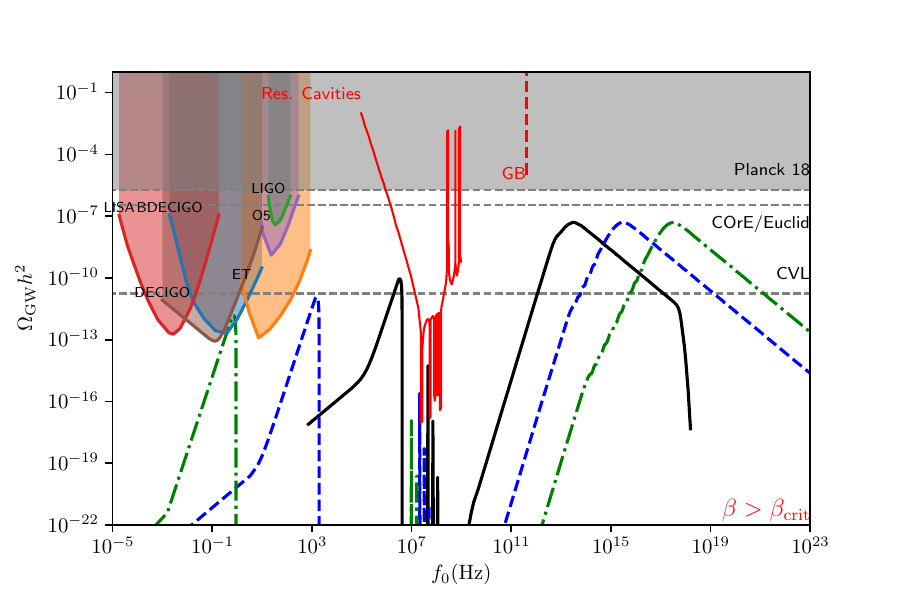}}
\subfigure{\includegraphics[width=\textwidth,height = 10cm]{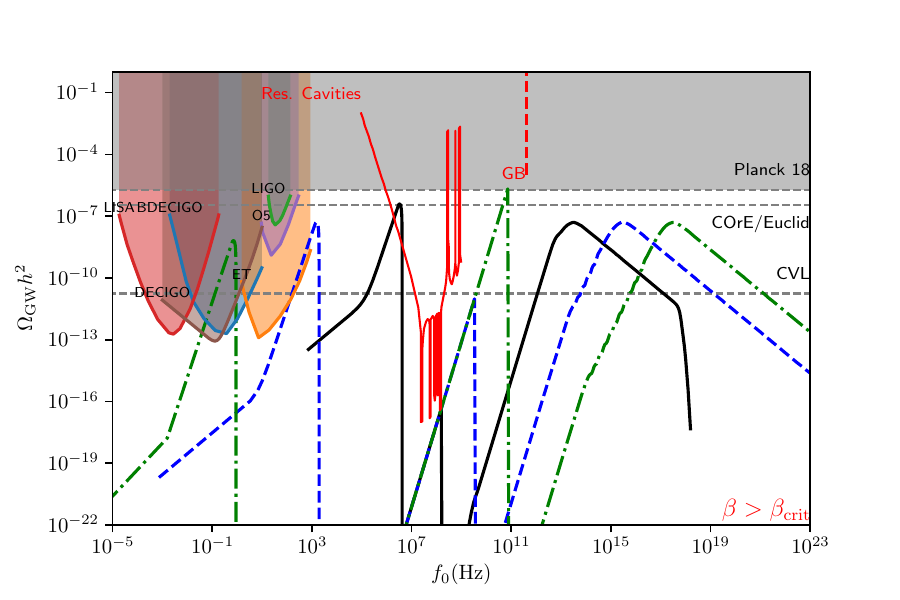}}
     \caption{Gravitational wave spectrum for $w_{\phi}=1/3$ (top) and $w_{\phi}=1$ (bottom) , considering an early matter-dominated phase with PBH domination ($\beta > \beta_c$). We show contributions from three sources: PBH evaporation (high-frequency peak), inflaton scatterings (mid-frequency peak), and density fluctuations in the PBH distribution (low-frequency peak). We display three different values for the PBH mass  $M_{\rm in}= 1 \, \rm{g}$ (black), $10^4 \, \rm{g} $ (blue) and $10^8 \, \rm{g} $ (green) for $\beta=10^{-4}$, $10^{-7}$ and $10^{-10}$ (top) and $\beta=10^{-7}$, $10^{-13}$ and $10^{-19}$(bottom), respectively. We also show the sensitivity curves for several future GW detectors: interferometers  (LISA, BDECIGO, ET, and LIGO), resonant cavities and gaussian-beam-based experiments. We also display current and projected experimental constraints from CMB measurements (Planck, COrE/Euclid and CVL).}
    \label{Fig:master1}
\end{figure*}

We summarize our results in Fig.~(\ref{Fig:master1})
. We considered an early matter domination period where PBHs dominate, i.e.~$\beta > \beta_c$, and combined the three sources of GWs: PBH evaporation, inflaton scatterings and density fluctuations in the PBH distribution. For $w_{\phi}=\frac{1}{3}$ and $1$, we display three different values for the PBH mass  $M_{\rm in}= 1 \, \rm{g}$ (black), $10^4 \, \rm{g} $ (blue) and $10^8 \, \rm{g} $ (green) for $\beta=10^{-4}$, $10^{-7}$ and $10^{-10}$, respectively. Apart from the resonant cavities and the GB-based experiments, we also included the sensitivity curves for future GW interferometers such as LISA \cite{Caprini_2019}, BDECIGO~\cite{Isoyama_2018,Breitbach_2019}, ET~\cite{Hild_2011}, LIGO O3~\cite{Abbott_2021}, LIGO O4 \cite{Kiendrebeogo:2023hzf} and LIGO O5 \cite{Kiendrebeogo:2023hzf}. The peak located at the higher end of the frequency spectrum corresponds to PBH evaporation, the one in the middle to inflaton scattering and the one at lower frequencies to density fluctuations in the PBH distribution. 

In the bottom panel of Fig.~(\ref{Fig:master1}), we see an enhancement in the peak of the GW spectrum obtained from the inflaton scattering. This effect arises from the redshift history induced by the PBH domination period. Unlike the usual reheating scenarios (inflaton decay), our scenario also depends on the PBH parameters ($M_{\rm{in}}$, $\beta$). For a stiff equation of state and a low $\beta$, the rapid decrease in the inflaton condensate energy density reduces the overall redshift experienced by the gravitons produced by scattering, resulting in an enhanced peak.

\begin{figure}[H]
    \centering
    \includegraphics[width=\linewidth]{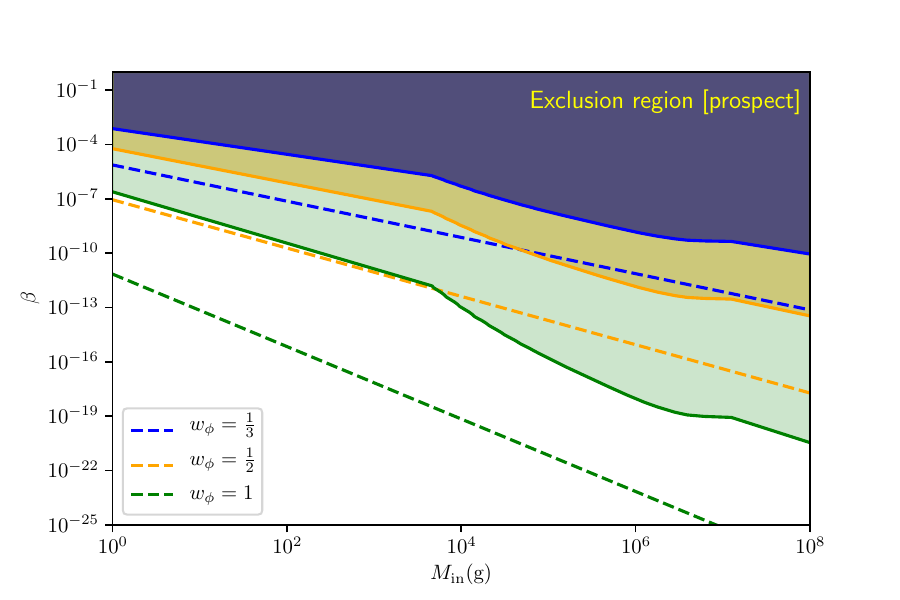}
    \caption{$\beta$ as a function of $M_{\rm in}$ for different values of $w_\phi=\frac13$ (blue), 
    $\frac12$ (orange) and 1 (green) here, the constraint assume that the next generation of detector will not see the isocurvature peak.
    }
    \label{Fig:newbetacrit}
\end{figure}

Finally, we show the prospects for the allowed parameter space in the ($\Min$, $\beta$) plane, considering the constraints from future GW experiments. We assumed that no detections would be made by the GW interferometers shown in Fig.~(\ref{Fig:newbetacrit}) within their entire sensitivity range. The testable parameter space with these projections is illustrated in Fig.~(\ref{Fig:newbetacrit}).

%%%%%%%%%%%%
\vspace{.2cm}
\section{Conclusion}\label{sec:concl}

In this paper we derived the GW spectrum produced by the evaporation of PBHs in a non-standard cosmology. We considered a Universe dominated by an 
inflaton condensate with an equation of state $P=w_\phi \rho$ at the end of inflation, whose oscillation sources GWs. Finally, we added the indirect GW 
contribution from the isocurvature perturbations produced by the decay of 
PBHs in the case where they eventually dominate the energy density of the 
Universe ($\beta > \beta_c$). Our results are summarized in Fig.~(\ref{Fig:master1}) (top), for $w_\phi=\frac13$, and Fig.~(\ref{Fig:master1}) (bottom), for $w_\phi=1$. We  
identify three distinct peaks arising from the different sources:  isocurvature perturbation, inflaton oscillations and PBH evaporation (from 
left to right). We also show current and projected experimental constraints from CMB measurements (Planck, COrE/Euclid and CVL) and projections for 
different GW detectors. Our findings highlight the importance of a coherent treatment of any process involving a population of PBHs in a non-standard 
expanding background, as they are intertwined to mutually influence the prediction, both in terms of the frequency of the signal and its amplitude. We found 
that even if the PBHs never come to dominate the energy density ($\beta < \beta_c$) and the lower-frequency peak is absent, our study remains valuable for future detections. The ratio of the remaining 
peak amplitudes from black hole evaporation and inflaton scattering is independent of the reheating temperature. Thus, detecting both peaks is 
crucial for identifying the common origin of the GWs: a population of primordial black holes in an inflation-dominated Universe.
Including and coherently combining the GW bremsstrahlung from the inflaton~\cite{Barman:2023ymn,Bernal:2023wus,Xu:2024fjl,Inui:2024wgj} or from massive particles produced by PBHs~\cite{Choi:2024acs} is another step, which is model-dependent, as the source of the bremsstrahlung must be defined.

%%%%%%%%%%%%
\section*{Acknowledgements}

This project has received support from the European Union's Horizon 2020 research and innovation program under the Marie Sklodowska-Curie grant agreement No 860881-HIDDeN and the CNRS-IRP project UCMN. The authors would like to acknowledge Guillem Domènech, Jan Tränkle, Simon Clery, and Jong-Hyun Yoon, for extremely useful discussions. MOOR is supported by the STFC under grant ST/X000753/1. RR acknowledges financial support from the STFC Consolidated Grant ST/X000583/1.

\bibliography{ref}

\clearpage

\end{document}